\documentclass[aps,prd,preprint,
superscriptaddress,preprintnumbers,eqsecnum,
showpacs,nofootinbib,nobibnotes]{revtex4}
\usepackage{amsfonts,amssymb,amsmath,bm}
\usepackage{graphicx}

\newcommand{\be}{\begin{equation}}
\newcommand{\bea}{\begin{eqnarray}}
\newcommand{\ee}{\end{equation}}
\newcommand{\eea}{\end{eqnarray}}
\newcommand{\bpi}{\begin{picture}}
\newcommand{\bce}{\begin{center}}
\newcommand{\epi}{\end{picture}}
\newcommand{\ece}{\end{center}}

\newcommand{\D}{\displaystyle}
\def\chic#1{{\scriptscriptstyle #1}}

\def\gb{{\Gamma}}
\def\g{\widetilde{\gb}}

\begin{document}

\title{Gluon mass generation without seagull divergences}

\author{Arlene~C. Aguilar}
\affiliation{Federal University of ABC, CCNH, 
Rua Santa Ad\'elia 166,  CEP 09210-170, Santo Andr\'e, Brazil.}

\author{Joannis Papavassiliou}
\affiliation{Department of Theoretical Physics and IFIC, 
University of Valencia-CSIC,
E-46100, Valencia, Spain.}

\begin{abstract}

Dynamical gluon  mass generation  has been traditionally  plagued with
seagull divergences, and all  regularization procedures proposed over the
years yield finite but scheme-dependent gluon masses.  In this work we
show how such divergences can be eliminated completely by virtue of a
characteristic  identity, valid  in  dimensional regularization.   The
ability to trigger the aforementioned identity hinges crucially on the
particular Ansatz  employed for  the three-gluon vertex  entering into
the Schwinger-Dyson equation governing  the gluon propagator.  The use
of  the  appropriate three-gluon  vertex  brings  about an  additional
advantage: one obtains two  separate (but coupled) integral equations,
one for the effective charge and  one for the gluon mass.  This system
of  integral  equations has  a  unique  solution, which  unambiguously
determines these  two quantities.  Most notably,  the effective charge
freezes  in  the  infrared,  and   the  gluon  mass  displays  
power-law  running in  the ultraviolet, in  agreement  with  earlier
considerations.

\end{abstract}


\maketitle

\section{Introduction}

The dynamical generation of a non-perturbative gluon mass has been first 
proposed by Cornwall~\cite{Cornwall:1982zr}, and has received significant attention 
over the years, both from the theoretical and the phenomenological point of view 
(see, e.g.,~\cite{Parisi:1980jy}).
According to this picture, even though  the  gluon is
massless  at the  level  of the  fundamental  Lagrangian, and  remains
massless to all order in perturbation theory, the non-perturbative QCD
dynamics  generate  an  effective,  momentum-dependent  mass,  without
affecting    the   local    $SU(3)_c$   invariance,    which   remains
intact.    The  generation of such a  mass  has been established
by studying the Schwinger-Dyson equations (SDEs)~\cite{Dyson:1949ha,Schwinger:1951ex} of QCD, in
a  gauge-invariant framework based on the  pinch technique  
(PT)~\cite{Cornwall:1982zr,Cornwall:1989gv,Binosi:2002ez,Nair:2005iw,Binosi:2009qm}, and its profound  
correspondence with the background field method (BFM)~\cite{Abbott:1980hw}. 

Specifically, when studying the SDE for the PT-BFM gluon propagator, $\Delta(q^2)$,  
one looks for infrared finite solutions, i.e. with  $\Delta^{-1}(0) > 0$ 
(see, e.g.,\cite{Cornwall:1982zr,Aguilar:2006gr,Aguilar:2007ie,Aguilar:2008xm}). 
Such solutions may  
be  fitted  by     ``massive''  propagators  of   the form \mbox{$\Delta^{-1}(q^2)  =  q^2  +  m^2(q^2)$}, where  
$m^2(q^2)$ is  not ``hard'', but depends non-trivially  on the momentum  transfer $q^2$.
In order to obtain  massive solutions gauge-invariantly, 
it  is necessary to invoke the well-known Schwinger mechanism~\cite{Schwinger:1962tn,Schwinger:1962tp}. 
In particular, one assumes that the strong QCD dynamics give rise to 
longitudinally-coupled 
composite (bound-state) massless poles~\cite{Jackiw:1973tr,Cornwall:1973ts,Eichten:1974et,Jackiw:1973ha,Farhi:1982vt,Frampton:2008zz}. 
These poles play  a  role rather  like Goldstone  excitations, in the sense that 
they preserve the form of the 
Ward identities satisfied by the Green's functions of the theory 
in the presence of a mass, but they are not associated with the breaking of any 
local or global symmetry.

When the  renormalization-group 
logarithms are  properly taken into  account in the SDE analysis, one obtains, in addition,  
the  non-perturbative  generalization  of  $\alpha(q^2)$, 
the  QCD  running  coupling (effective charge), of the form~\cite{Cornwall:1982zr,Aguilar:2008fh,Aguilar:2009nf}
$\alpha^{-1}(q^2)= b \ln\left(\frac{q^2+\,4\, m^2(q^2)}{\Lambda^2}\right)$. 
The presence of $m^2(q^2)$ in the argument of the logarithm 
tames  the   Landau  singularity   associated   with  the
perturbative $\beta$  function, and the resulting  effective charge is
asymptotically free in  the ultraviolet , ``freezing'' at a  finite value in the infrared, namely  
$\alpha^{-1}(0)= b \ln (4m^2(0)/\Lambda^2)$.

As has been emphasized in the literature~\cite{Cornwall:1979hz,Cornwall:1997ds},
the generation of a gluon mass 
is intimately connected with 
a variety of other related phenomena, and most importantly with  
the center vortex picture of confinement~\cite{de Forcrand:1999ms,Greensite:2003bk,Greensite:2006sm}. 
In particular, an effective low-energy field 
theory for describing the gluon mass   
is  the  gauged non-linear sigma model  known  as ``massive
gauge-invariant Yang-Mills''~\cite{Cornwall:1979hz}. 
 This model admits  vortex
solutions,  with a  long-range pure  gauge term  in  their potentials,
which endows  them with a topological quantum  number corresponding to
the center  of the gauge group  [$Z_N$ for $SU(N)$], and  is, in turn,
responsible for quark  confinement and gluon screening~\cite{Cornwall:1997ds} 
Specifically, center vortices of  thickness $\sim m^{-1}$ are assumed to 
form a condensate because their entropy
(per  unit  size) is  larger  than  their  action.  This  condensation
furnishes an  area law to  the fundamental representation  
Wilson loop, thus confining quarks~\cite{Brodsky:2008be}.

The general picture described above appears to be in qualitative agreement with 
a plethora of lattice simulations, where  
the gluon propagators (in various gauges) reach a finite (non-vanishing) value in the deep 
infrared, as would happen in the presence of a ``mass''~\cite{grilatt}.  
This  rather characteristic  behavior 
was already observed in early studies~\cite{Alexandrou:2000ja}, and 
has been  firmly established  recently (in the Landau gauge) using
large-volume lattices, for both $SU(2)$~\cite{Cucchieri:2007md} and $SU(3)$~\cite{Bogolubsky:2007ud,Bowman:2007du,Iritani:2009mp} 
pure  Yang-Mills (no quarks included). 

It is also important to mention that a qualitatively similar situation emerges within the 
``refined'' Gribov-Zwanziger formalism, presented in~\cite{Dudal:2008sp}. 
In this latter framework the gluon mass is obtained   
through the addition of appropriate condensates 
to the original Gribov-Zwanziger action~\cite{Gribov:1977wm,Zwanziger:1993dh}. 
Interestingly enough, one obtains a gluon mass displaying power-law running, in     
agreement with a variety of independent studies~\cite{Lavelle:1991ve,Aguilar:2007ie}), 
as well as the results of the present article, as explained below.

Since the dynamical generation of an effective gluon mass is a purely non-perturbative effect,   
its technical implementation is rather intricate, and requires the harmonious synthesis  
of several ingredients~\cite{Jackiw:1973tr,Cornwall:1973ts,Eichten:1974et,Jackiw:1973ha,Farhi:1982vt,Frampton:2008zz}. 
In particular, the exact way how the  
Schwinger's mechanism will be employed 
is crucial for the self-consistency of the entire picture.  
Turns out that one characteristic drawback 
in the realization of this dynamical scenario   
can in fact be traced back to a certain subtlety in  the implementation of the Schwinger's mechanism
at the level of the relevant SDE.   

Specifically, the massless poles necessary for triggering Schwinger's mechanism and  
allowing the possibility of a non-vanishing $\Delta^{-1}(0)$, 
enter into the SDE for the gluon propagator through the particular 
Ansatz employed for the fully-dressed three-gluon vertex. 
Of course, a physically motivated  Ansatz  
must satisfy, in addition, the correct WI, in order to preserve the 
transversality of the gluon self-energy. 
Even though several such  Ans\"atze have been proposed 
over the years~\cite{Cornwall:1982zr,Aguilar:2006gr,Aguilar:2007ie,Aguilar:2008xm}, they all suffer 
from a typical problem: as desired,  $\Delta^{-1}(0)$ 
does not vanish; however, its value is expressed in terms of seagull integrals, i.e. 
divergent integrals of the type $\int_k \Delta(k)$ and $\int_k k^2 \Delta^2(k)$.  
This fact, in turn, introduces the need to make sense out of these divergences,
given that one is not allowed to
absorb them into a counterterm of the type $ m^2_0 (\Lambda^2_{\chic{\mathrm{UV}}}) A^2_{\mu}$, 
because this would compromise the gauge invariance of the original Lagrangian, 
which at no point is to be modified. Even though a variety of 
regularizations have been proposed in the literature cited above, 
it is clear that the appearance of these divergences, the need to regularize them, and the 
ambiguities resulting in from such a regularization, are without a doubt some 
of the weakest theoretical points of this entire construction.

In this paper we present  
a more refined Ansatz for the three-gluon vertex, 
which completely eliminates all seagull divergences.
This new Ansatz is inspired from the 
photon-scalar vertex of {\it  scalar} QED, 
introduced by  Ball and Chiu~\cite{Ball:1980ay}.  When inserted into the 
gluon SDE obtained within the PT-BFM formalism, 
leads to the elimination of all seagull divergences, by triggering  
a special identity, valid in dimensional regularization (DR), yielding finally  
a non-vanishing and {\it finite} value for $\Delta^{-1}(0)$. 

In the context of scalar QED, 
the identity in question, given in Eq.~({\ref{basid}}), is instrumental  
in enforcing the masslessness of the photon, in the absence of 
any bound-state poles, i.e. when the Schwinger mechanism is not in operation. 
Specifically, the aforementioned  Ansatz of ~\cite{Ball:1980ay}, when incorporated into the 
SDE for the photon, gives rise to a $\Delta^{-1}(0)$ that is expressed 
in terms of seagull contributions, which do not vanish individually, due to the 
simple fact that the charged scalars are massive already at tree level.
However, the vertex of Ball and Chiu is such that the divergent seagull terms appear precisely in the unique 
combination that will lead to their mutual annihilation, due to the identity of Eq.~({\ref{basid}}).

The proposed three-gluon vertex 
consists of two parts: {\it (i)}
a part that leads to the cancellation of all seagull divergences by virtue of  the identity of 
Eq.~({\ref{basid}}), exactly as happens in the scalar QED case; the only difference
is that now the seagull terms in question 
originate from the  gluonic self-interactions, i.e. they are  
composed by the (effectively massive) gluon propagator.
{\it (ii)} a part that contains massless bound-state poles, 
thus enforcing the Schwinger mechanism. It is from this second part of the vertex 
that, after solving the resulting integral equation, one 
finally obtains a finite value for  $\Delta^{-1}(0)$. 

In addition to eliminating the seagull divergences, the use of the aforementioned 
vertex brings about a further important advantage. Specifically, 
the SDE for the gluon propagator of the PT-BFM may be separated unambiguously into two 
distinct but coupled integral equations, one governing the evolution of the 
effective charge (running coupling), ${\overline g}^2(q^2)$, 
and one determining the momentum-dependence of the 
effective gluon mass, $m^2(q^2)$.
This is to be contrasted with the standard procedure followed in the 
literature, where the SDE equation is solved for the renormalization-group (RG) invariant 
combination ${\widehat d}(q^2) = g^2 \Delta (q^2)$, which is subsequently decomposed 
into an effective charge  and a running mass 
according to  \mbox{${\widehat d}(q^2) = {\overline g}^2(q^2)/(q^2+m^2(q^2))$},  
by imposing 
physically motivated constraints on the form of ${\overline g}^2(q^2)$ and $m^2(q^2)$.
This procedure 
suffers from the 
obvious ambiguity of trying to extract two components out of a given function; instead,  
the new procedure, involving two individual equations, furnishes uniquely  ${\overline g}^2(q^2)$ and $m^2(q^2)$, 
and it is the ${\widehat d}(q^2)$ that is subsequently obtained uniquely, by combining these two quantities.

The present article is organized as follows. 
In Section~\ref{mgsd} we review the salient features of dynamical 
gauge-boson mass generation through the  Schwinger mechanism, 
which constitutes the cornerstone of the entire approach. We explain how the 
aforementioned mechanism must be judiciously incorporated into the SDE equations
of QCD, and the crucial role played by the three-gluon vertex. The problem of the seagull divergences, 
which is endemic to all existing approaches, is discussed, and some examples 
of (not fully satisfactory) attempts for its resolution are mentioned.
In Section~\ref{sqed} we turn to the instructive case of scalar QED, 
and demonstrate in detail how the seagull divergences cancel out from 
the SDE for the photon propagator, by virtue of the identity of Eq.~({\ref{basid}}),  
which is in turn triggered by the vertex Ansatz of ~\cite{Ball:1980ay}. A counter-example of a vertex that 
does not trigger the identity is also discussed. 
In Section~\ref{fgmg} we apply the lessons of the previous section to the 
case of (quarkless) QCD. In particular, an improved Ansatz for the three-gluon vertex is 
constructed, which incorporates the Schwinger mechanism through the appearance of massless poles, and, at the 
same time, triggers the identity of Eq.~({\ref{basid}}), leading to total seagull annihilation. 
In Section~\ref{coupeq} we obtain the system of two coupled integral equations that  
determine the momentum dependence of two RG-invariant quantities, 
namely ${\overline g}^2(q^2)$ and $m^2(q^2)$, for the 
{\it entire} range of physical momenta, i.e. from the deep IR to the deep UV.    
The system is solved numerically and the obtained solutions are discussed. 
Most notably, $m^2(q^2)$ display power-law running, in agreement with various earlier considerations. 
Finally, in Section~\ref{concl} we summarize our conclusions. 
In addition, in three Appendices we derive in detail various intermediate results used throughout the article. 

\section{\label{mgsd}Mass generation and the problem of seagull divergences.}

The gluon propagator, $\Delta_{\mu\nu}(q)$, in covariant gauges (in particular,
linear, $R_{\xi}$-type of gauges, and the BFM) has the form  
\be 
\Delta_{\mu\nu}(q)= -i\left[ {\rm P}_{\mu\nu}(q)\Delta(q^2) +\xi\frac{\D q_\mu
q_\nu}{\D q^4}\right] \,,
\label{fprop}
\ee
where $\xi$ denotes the gauge-fixing parameter, and 
the  transverse projector  ${\rm P}_{\mu\nu}(q)$ is given by 
\be
{\rm P}_{\mu\nu}(q)= g_{\mu\nu} - \frac{q_\mu q_\nu}{q^2} \,.
\label{proj}
\ee

The scalar factor $\Delta(q^2)$ is given by 
\be
\Delta^{-1}(q^2) = q^2 + i \Pi(q^2)\,,
\label{func}
\ee 
where  $\Pi_{\mu\nu}(q)={\rm P}_{\mu\nu}(q) \,\Pi(q^2)$ is the gluon self-energy.  
One usually defines 
the dimensionless vacuum polarization, to be denoted by ${\bf \Pi}(q^2)$, 
as 
$\Pi(q^2)=q^2 {\bf \Pi}(q^2)$, and thus  
\be
\Delta^{-1}(q^2) = q^2 [1 + {\bf \Pi}(q^2)]\,.
\label{funcbar}
\ee

As Schwinger pointed out long time ago~\cite{Schwinger:1962tn},     
the gauge invariance of a vector field does not necessarily 
imply zero mass for the associated particle, if the 
current vector coupling is sufficiently strong. 
Schwinger's fundamental observation 
was that if (for some reason) the vacuum polarization 
of the gauge bosons acquires a pole at zero momentum transfer, then the 
 vector meson becomes massive, even if the gauge symmetry 
forbids a mass at the level of the 
fundamental Lagrangian \cite{Schwinger:1962tp}.
Indeed, casting the self-energy in the form of (\ref{funcbar}), 
it is clear that if ${\bf \Pi}(q^2)$ has a pole at  $q^2=0$ with positive 
residue $\mu^2$, i.e. ${\bf \Pi}(q^2) = \mu^2/q^2$, then (in Euclidean space)
\be
\Delta^{-1}(q^2) = q^2 + \mu^2\,.
\label{funcbar2}
\ee 
Thus, the vector meson 
becomes massive, $\Delta^{-1}(0) = \mu^2$, 
even though it is massless in the absence of interactions ($g=0$). 

There is {\it no} physical principle which would preclude ${\bf \Pi}(q^2)$ from 
acquiring a pole~\cite{SM}.  
Actually, the appearance of the required pole may happen for purely dynamical reasons, 
and, in particular, {\it without} the 
need to introduce fundamental scalar field in the Lagrangian~\cite{Higgs}.
Since bound states are expected to exist in most physical systems
one may suppose that, for sufficiently 
strong binding, the mass of such a bound state will be reduced to zero, thus generating a mass
for the vector meson without interfering with gauge 
invariance~\cite{Jackiw:1973tr,Cornwall:1973ts,Eichten:1974et,Jackiw:1973ha,Farhi:1982vt,Frampton:2008zz}.

When  applying   the  dynamical  concepts  described   above  to  pure
Yang-Mills theories,  such as  quarkless QCD, one  assumes that,  in a
strongly-coupled   gauge  theory   longitudinally   coupled  zero-mass
bound-state excitations are dynamically produced ~\cite{pogg}. Thus, it
is clear that a vital ingredient for this scenario is strong coupling,
which can only  come from the infrared instabilities  of a non-abelian
gauge  theories.   The   aforementioned  excitations  are  {\it  like}
dynamical Nambu-Goldstone bosons, in the sense that they are massless,
composite,  and longitudinally coupled;  but, at  the same  time, they
differ  from  Nambu-Goldstone  bosons   as  far  as  their  origin  is
concerned: they  do {\it not} originate from  the spontaneous breaking
of  any global symmetry.   The main  role of  these excitations  is to
trigger the Schwinger mechanism, i.e.  to provide the required pole in
the  gluon self-energy, and  more specifically,  the gauge-independent
${\bf \Pi}(q^2)$ obtained   with  the  PT,   thus  furnishing  a
gauge-independent dynamical mass for the gluons~\cite{polcan}.

Of course, in order to obtain the full dynamics, 
such as, for example, the momentum-dependence of the dynamical mass, 
one must turn eventually  to the SDE that governs 
the corresponding gauge-boson self-energy. 
The way the Schwinger mechanism
is integrated into the SDE 
is through the form of the three-gluon vertex.
The latter, even in the absence of mass generation, constitutes a  
central ingredient of the SDE, and plays a crucial role in 
enforcing the transversality of the gluon self-energy.    
Therefore, an important requirement for any self-consistent Ansatz used for that 
vertex is that it should satisfy the correct WI (or STI) of the PT-BFM formulation, namely  
\be
q^{\mu}{\g}_{\mu\alpha\beta}= \Delta^{-1}_{\alpha\beta}(k+q) -\Delta^{-1}_{\alpha\beta}(k)\,.
\label{VWIf}
\ee
In addition, in order to generate a dynamical mass  
one must assume that the vertex contains {\it dynamical poles}, which 
will trigger the Schwinger mechanism when 
inserted into the SDE for the gluon self-energy.

The point is that the full realization of the procedure outlined above  
is very subtle. 
In particular, even though the use of a three-gluon vertex 
containing massless poles and satisfying the correct WI
leads indeed 
to a transverse and infrared finite self-energy (i.e. $\Delta^{-1}(0) \neq 0$), 
as expected, 
the actual value of $\Delta^{-1}(0)$ has always been 
expressed in terms of divergent integrals, of the form 
(see, e.g.,\cite{Cornwall:1982zr,Aguilar:2006gr,Aguilar:2007ie,Aguilar:2008xm})
\be
\Delta^{-1}(0) =  c_1 \int_k \Delta(k) + c_2 \int_k k^2 \Delta^2(k) \,,
\label{dqd}
\ee
where (in DR)    
\mbox{$\int_{k}\equiv\mu^{2\varepsilon}(2\pi)^{-d}\int\!d^d k$}, 
with $d=4-\epsilon$ the dimension of space-time~\cite{qdi}.
This is not a problem, in principle, provided that the 
divergent integrals appearing on the rhs of (\ref{dqd})
can be properly regulated and made finite, {\it without} 
introducing counterterms of the 
form $ m^2_0 (\Lambda^2_{\chic{\mathrm{UV}}}) A^2_{\mu}$, 
which are forbidden by the local gauge invariance 
of the  fundamental  QCD Lagrangian.
However, various regularization procedures introduced in the literature  
have been eventually thwarted by 
all sorts of additional complications of variable severity. 

The simplest regularization possibility, for example, is to employ the usual DR trick for eliminating  
quadratic divergences, namely subtract  $\int_k k^{-2}=0$. 
Assuming a form \mbox{$\Delta(k) = k^2 + m^2(k)$}, this standard (and completely legitimate) operation, 
\be
\int_k \Delta(k)= \int_k \frac{1}{k^2 + m^2(k)} - \int_k \frac{1}{k^2}
= - \int_k \frac{m^2(k)}{k^2[k^2 + m^2(k)]}\,,
\ee
leads to a finite integral, provided $m^2(k)$ drops off sufficiently fast in the 
UV, a feature which is in any case expected from a dynamically generated mass. 
The general problem with this procedure, however, is the reversal of sign that it induces~\cite{Cornwall:1982zr}, 
which eventually clashes with the requirement of a positive-definite $\Delta^{-1}(0)$. 

In a recent work~\cite{Aguilar:2008xm} the aforementioned procedure was refined in such a way as to 
evade the sign problem. The general idea is to eliminate the perturbative 
tail of $\Delta(k)$ by subtracting out DR ``zeros'', using the 
generalized formula 
\be 
\int_k  \frac{\ln^{n}\! k^2}{k^2}=0,  \,\,\, n={0,1,2,}\dots
\label{dimreg}
\ee  
Specifically, for large enough $k^2$,   
$\Delta(k^2)$ goes over to its 
perturbative expression, to be denoted by 
$\Delta_{\rm pert}(k^2)$; it has the form  
\be
\Delta_{\rm pert}(k^2)=\sum_{n=0}^N a_n\frac{\ln^{n} k^2}{k^2}, 
\ee
where the coefficient $a_n$ are known from the perturbative 
expansion.
Then one may use (\ref{dimreg}) to regularize the rhs of (\ref{dqd}), and obtain   
\be
16 \pi^2 \Delta^{-1}_{\rm reg}(0) =
c_1\int_0^s\!\!dy\ y
\left[   \Delta(y)    -\Delta_{\rm pert}(y)\right] +   
c_2 \int_0^s\!\!dy\ y^2\left[\Delta^2(y)-\Delta^2_{\rm pert}(y)\right]\,,
\label{regtad}
\ee
which is finite (and has been shown to be positive). 
As explained in~\cite{Aguilar:2008xm}, the obvious ambiguity of this procedure is 
the choice of the point $s$, past which the two 
curves, $\Delta(y)$ and $\Delta_{\rm pert}(y)$, are 
assumed to coincide (and cancel exactly against each other). 
Thus, the actual value of $\Delta^{-1}_{\rm reg}(0)$ remains 
largely undetermined. 
Even though additional qualitative arguments may be used to restrict the 
allowed interval of $s$, thus  
reaching good agreement with recent lattice data,
from the theoretical point of view it is clear that this issue 
is far from settled.

\section{\label{sqed}Scalar QED and the seagull identity}

In this section we will study some of the basic issues related to the 
appearance and cancellation of seagull divergences
in the context of a theory much simpler than QCD, namely scalar QED. 
Specifically, we will study the SDE governing the photon, and we will 
discover a basic identity, which, in the absence of 
massless poles (i.e., with the Schwinger mechanism ``switched off'') 
enforces the masslessness of the photon, despite the fact that 
individual seagull contributions do not vanish.  
In addition, we will see through an explicit detailed construction 
that the Ans\"atze employed for 
the all-order photon-scalar vertex entering into the SDE are crucial for 
the activation of this identity. 

\begin{figure}[!t]
\begin{center}
\includegraphics[scale=0.7]{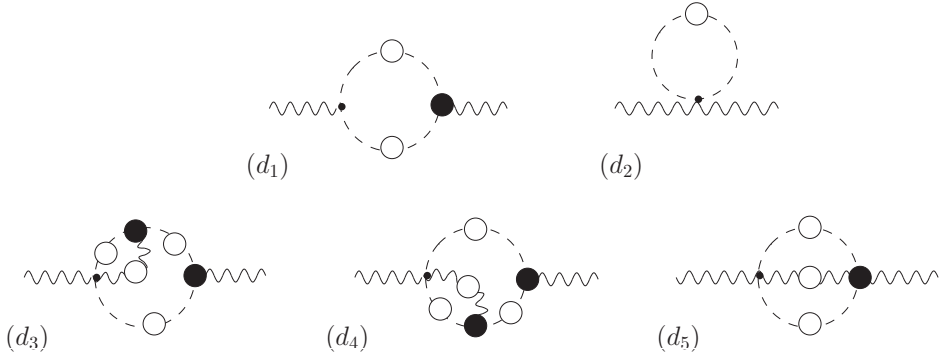}
\end{center}
\vspace{-0.5cm}
\caption{Diagrams contributing to the SDE for the photon self-energy in scalar QED.}
\label{p1}
\end{figure}

The SDE for the photon of scalar QED is shown in Fig.~\ref{p1}.
It is a straightforward exercise to demonstrate that, 
by virtue of the Abelian WI's satisfied by the full vertices of the theory, 
the SDE may be truncated ``loop-wise'', without compromising the transversality of the photon, i.e., 
$q^{\mu} \Pi_{\mu\nu}^{[(d_1)+(d_2)]} = q^{\mu} \Pi_{\mu\nu}^{[(d_3)+(d_4)+(d_5) ]} = 0$ . 

\begin{figure}[!t]
\begin{center}
\includegraphics[scale=0.7]{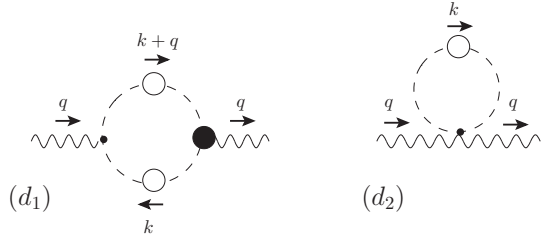}
\end{center}
\vspace{-1.0cm}
\caption{The ``one-loop dressed'' SDE for the photon self-energy.}
\label{p2}
\end{figure}

At the ``one-loop dressed'' level the SDE for the photon self-energy reads (Fig.~\ref{p2}) 
\be
\Pi_{\mu\nu}(q)= e^2 \int_k \Gamma_{\mu}^{(0)}{\mathcal D}(k){\mathcal D}(k+q)\gb_{\nu}
  + e^2 \int_k \Gamma_{\mu\nu}^{(0)}{\mathcal D}(k) \,,
\label{sself}
\ee
where ${\mathcal D}(k)$  is the fully-dressed propagator of the scalar field.
$\gb_{\nu}$ is the fully dressed photon-scalar vertex, whose 
tree-level expression is given by $\Gamma_{\mu}^{(0)}= -i(2k+q)_{\mu}$.  
Moreover, the bare quadrilinear photon-scalar vertex 
is given by $\Gamma_{\mu\nu}^{(0)}= 2ig_{\mu\nu}$. 
The  photon-scalar vertex $\Gamma_{\mu}$ and the scalar propagator  ${\mathcal D}$ 
are related by the Abelian all-order WI 
\be
q^{\nu}\gb_{\nu}= {\mathcal D}^{-1}(k+q) -{\mathcal D}^{-1}(k) \,.
\label{sward}
\ee
It is fairly easy  to demonstrate the transversality of $\Pi_{\mu\nu}(q)$, namely  
that $q^{\nu} \Pi_{\mu\nu}(q)=0$. 
To that end, act with $q^{\nu}$ on the two terms on the rhs of (\ref{sself}), 
 and notice that, by virtue of (\ref{sward}), the (contracted) first term 
(after appropriate shifting of the integration variable, 
a legitimate operation in DR), cancels exactly against the second. 
Given that $\Pi_{\mu\nu}(q)$ is transverse, 
it assumes the form 
$\Pi_{\mu\nu}(q)= \Pi(q^2){\rm P}_{\mu\nu}(q)$; thus
one may determine $\Pi(q^2)$ simply by taking the trace 
of both sides of (\ref{sself}), i.e. 
\be
\Pi(q^2) = \frac{-2ie^2}{d-1}
\left[ \int_k {\mathcal D}(k){\mathcal D}(k+q) k^{\mu} \gb_{\mu} - d  \int_k  {\mathcal D}(k) \right]\,,
\label{scalpi}
\ee
where Eq.~(\ref{sward}) was used.

Let us compute from (\ref{scalpi}) the one-loop expression for $\Pi(q^2)$, to be denoted by $\Pi^{(1)}(q^2)$.
We have (we are using DR throughout)
\be
\Pi^{(1)}(q^2) = \frac{-ie^2}{d-1}
\left[ \int_k (4k^2- q^2){\mathcal D}_0(k){\mathcal D}_0(k+q)  - 2d  \int_k  {\mathcal D}_0(k) \right]\,,
\label{scalpi1}
\ee
where ${\mathcal D}_0(k) = (k^2-m^2)^{-1}$. Taking the limit $q\to 0$, we find  
\be
\Pi^{(1)}(0) = \frac{-4ie^2}{d-1}
\left[ \int_k k^2 {\mathcal D}_0^2(k)  - \frac{d}{2}  \int_k  {\mathcal D}_0(k) \right]\,.
\label{scalpi10}
\ee 
Of course, there is no doubt that the 
photon remains massless perturbatively, i.e. we must have that $\Pi^{(1)}(0)=0$. 
However, the way this requirement is realized is rather subtle:
the rhs of (\ref{scalpi10}) vanishes indeed, by virtue of an identity 
that is exact in DR, namely 
\be
\int_k \frac{k^2}{(k^2-m^2)^2} = \frac{d}{2} \int_k \frac{1}{k^2-m^2}\,,
\label{id1}
\ee
or, equivalently, 
\be
2 m^2\int_k \frac{1}{(k^2-m^2)^2} = (d-2)\int_k \frac{1}{k^2-m^2}\,.
\label{id2}
\ee
The relations given in (\ref{id1}) and (\ref{id2}) can be easily verified using the standard 
integration rules of the DR~\cite{Peskin:1995ev}. Thus, the perturbative masslessness of the photon is 
explicitly realized and self-consistently enforced within the DR.
Note that Eq.(\ref{id1}) may be cast 
in a form that is particularly suggestive for the analysis that follows, 
namely 
\be
\int_k \,k^2\frac{\partial {\mathcal D}_0(k) }{\partial k^2}= 
- \frac{d}{2} \int_k {\mathcal D}_0(k)\,.
\label{basid0}
\ee
To demonstrate (\ref{basid0}) directly, i.e. without deducing it from (\ref{id1}), 
we first go to Euclidean space, 
use spherical coordinates (see \ref{spher}), and integrate by parts (in $d$ dimensions).
Setting $k^2_E = y$, we have (suppressing the angular integral)
\be
\int_0^{\infty} dy\, y^{\frac{d}{2}}\,\frac{\partial {\mathcal D}_0(y)}{\partial y}= 
\left[y^{\frac{d}{2}}{\mathcal D}_0(y)\right]^{\infty}_0
- \frac{d}{2}  \int_0^{\infty} dy \,y^{(\frac{d}{2}-1)} {\mathcal D}_0(y)\,.
\label{partint}
\ee
Evidently, dropping the surface term, an operation that can be formally 
justified by the standard analytic continuation employed within the DR (see, e.g.~\cite{Collins:1984xc}), 
yields immediately Eq.~(\ref{basid0}).

We now return to the general Eq.(\ref{scalpi}). In order to analyze it further 
we must furnish some information about the form of $\gb_{\mu}$. Of course, any meaningful Ansatz for 
$\gb_{\mu}$ must satisfy the WI of (\ref{sward}), or else the transversality of 
$\Pi_{\mu\nu}(q)$ will be compromised from the outset. 
The form obtained by Ball and Chiu \cite{Ball:1980ay}, 
after ``solving'' the WI, under the additional  
requirement of not introducing kinematic singularities, is   
\be
\gb_{\mu}= \frac{(2k+q)_{\mu}}{(k+q)^2-k^2}\left[{\mathcal D}^{-1}(k+q) -{\mathcal D}^{-1}(k)\right]
+ A(k,q)\left[(k+q)\cdot q \,k_{\mu} -k\cdot q\, (k+q)_{\mu}\right]\,,
\label{strans_vert}
\ee
where $A(k,q)$ is finite as $q\to 0$. Clearly the first term 
satisfies (\ref{sward}), while 
the part proportional to $A(k,q)$ is identically conserved.
 
It is easy to recognize that when this latter term is inserted  
into (\ref{scalpi}) it yields a contribution that vanishes as $q\to 0$,
with no additional assumptions, other than the regular nature of $A(k,q)$; 
we will therefore neglect that term in what follows. On the other hand, 
the first term of $\gb_{\mu}$ yields 
\be
\Pi(q^2) = \frac{ie^2}{d-1}
\left[ \int_k (4k^2- q^2)\frac{{\mathcal D}(k+q)-{\mathcal D}(k)}{(k+q)^2-k^2} +  
2d  \int_k  {\mathcal D}(k) \right]\,,
\label{scalnp}
\ee
Taking the limit of Eq.(\ref{scalnp}) as \mbox{$q\to 0$}, using that 
\be
\frac{{\mathcal D}(k+q)-{\mathcal D}(k)}{(k+q)^2-k^2} \to \frac{\partial {\mathcal D}(k)}{\partial k^2} + 
{\cal O}(q^2) \,,
\ee
we have that  
\be
\Pi(0)= \frac{4ie^2}{d-1}\left[
\int_k\, k^2\frac{\partial {\mathcal D}(k)}{\partial k^2} + \frac{d}{2}\,\int_k {\mathcal D}(k)\right],
\label{sself1}
\ee
Of course, we must have that $\Pi(0)=0$, given that there is nothing 
in the dynamics that could possibly endow the photon with a mass; in particular, we have 
not employed Schwinger's mechanism, i.e. we have not introduced dynamical poles, 
and, given the form of (\ref{strans_vert}), 
neither kinematic ones, which might simulate the dynamical ones at the level of the SDE (see below).
Thus, the rhs of (\ref{sself1}) must vanish, and therefore, we must have that 
\be
\int_k \,k^2 \frac{\partial {\mathcal D}(k)}{\partial k^2}= - \frac{d}{2}\int_k\, {\mathcal D}(k) \,,
\label{basid}
\ee
which is the non-perturbative generalization of (\ref{basid0}); its demonstration 
proceeds exactly in the same way (and under the same assumptions).

Note a crucial point: the seagull terms appearing in (\ref{basid0})
{\it cannot} be set to zero individually, 
because the scalar propagator inside them is massive (already at tree-level):
the only way to keep the photon massless, is to employ (\ref{basid0}), 
which cancels them against each other. For example, if the 
term $\int_k {\mathcal D}(k)$ on the rhs were multiplied by any factor other than 
$(d/2)$ one would be stuck with seagull divergences.

Let us now try a different Ansatz for $\gb_{\mu}$, which, due to its special form 
will not trigger Eq.~(\ref{basid}),
and thus will lead to a non-vanishing (but divergent) value for $\Pi(0)$. Specifically, consider the 
vertex given by 
\be
\gb_{\mu}= \Gamma_{\mu}^{(0)} + 
\frac{q_{\mu}}{q^2}\left[ \Sigma(k+q) - \Sigma(k)\right]\,,
\label{vertpol1}
\ee
where $\Sigma(k)$ is the self-energy of the scalar field, ${\mathcal D}^{-1}(k) = k^2 - m^2 + \Sigma(k)$.  
Equivalently, we may write  
\be
\gb_{\mu}= \left\{\Gamma_{\mu}^{(0)} - \frac{q_{\mu}}{q^2}[(k+q)^2 - k^2]\right\}
+ \frac{q_{\mu}}{q^2}\left[{\mathcal D}^{-1}(k+q) -{\mathcal D}^{-1}(k)\right]\,.
\label{vertpol2}
\ee
The $\gb_{\mu}$ in (\ref{vertpol1})-(\ref{vertpol2})
satisfies again the WI of (\ref{sward}), and thus, as before, 
the transversality of the vacuum polarization is guaranteed. 
There is an important difference, however, 
between (\ref{strans_vert}) and (\ref{vertpol2}):  the latter contains 
massless poles, and thus, is capable of giving rise to a non-vanishing $\Delta^{-1}(0)$. 

Indeed, substituting 
$\gb_{\mu}$ of Eq.~(\ref{vertpol2}) 
into (\ref{scalpi}), after straightforward algebra we obtain 
\be
\Pi(q^2)= \frac{4ie^2}{d-1}\left[
\int_k\, {\mathcal D}(k) {\mathcal D}(k)(k+q) \left[\frac{(k \cdot q)^2}{q^2} - k^2\right] 
+ \frac{d-1}{2}\,\int_k {\mathcal D}(k)\right],
\label{sselfwp}
\ee
Using that 
\be
\int_k \frac{(k \cdot q)^2}{q^2} {\mathcal D}(k) {\mathcal D}(k+q)\bigg|_{q^2\to 0}
= \frac{1}{4} \int_k \,k^2 \,{\mathcal D}^2(k)\,,
\ee
we find from (\ref{sselfwp}) (setting $d=4$)
\be
\Pi(0)= ie^2 \left[ 2\int_k {\mathcal D}(k) - \int_k  k^2 {\mathcal D}^2(k)\right]\,,
\label{sselfwp0}
\ee
which has the general form given in (\ref{dqd}). 
Evidently, the Ansatz of (\ref{vertpol1}) does not trigger Eq.~(\ref{basid}), 
and one ends up with a non-zero $\Delta^{-1}(0)$, which, however, is expressed in terms 
of divergent seagull-type integrals. 

It is evident from the above analysis that the massless poles, indispensable as they may be 
for generating a non-vanishing $\Delta^{-1}(0)$, must be incorporated into the 
SDE with particular care, or else they give rise to seagull divergences. 
But even without this pathology, it is clear that the vertex of (\ref{vertpol1})
does not constitute an optimal Ansatz. For example,  
if the $(1/q^2)$ pole is considered to be of purely non-perturbative origin (as it is supposed to), 
it has vanishing perturbative expansion, and so, 
to all orders in perturbation theory $\gb_{\mu}=\Gamma_{\mu}^{(0)}$, which 
is of course not correct. 

In the next section 
we will see that the correct procedure is to add 
to the vertex of (\ref{strans_vert})  
non-perturbative pole terms, in such a way as to preserve the seagull cancellation
implemented by Eq.~(\ref{basid}), and, at the same time, obtain a finite  
$\Delta^{-1}(0)$.  


\section{\label{fgmg}Finite gluon mass generation}

\begin{figure}[!b]
\begin{tabular}{c}
\begin{minipage}[b]{0.25\linewidth}
\centering
\includegraphics[scale=0.6]{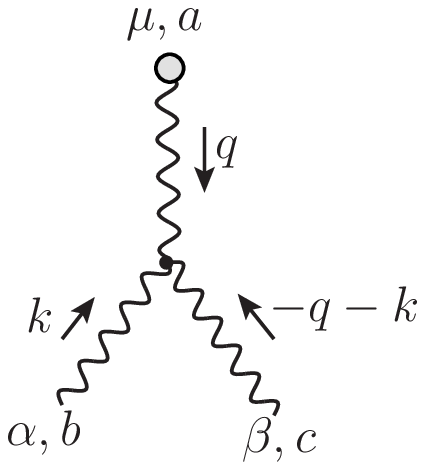}
\end{minipage}
\begin{minipage}[b]{0.75\linewidth}
\begin{center}
\bea
gf^{abc}\left[(2k+q)_{\mu}g_{\alpha\beta} + 2 q_{\beta}g_{\mu\alpha}- 2 q_{\alpha}g_{\mu\beta} \right] \,,  \\ \nonumber
 \nonumber
\eea
\end{center}
\end{minipage} \\
\begin{minipage}[b]{0.25\linewidth}
\centering
\includegraphics[scale=0.45]{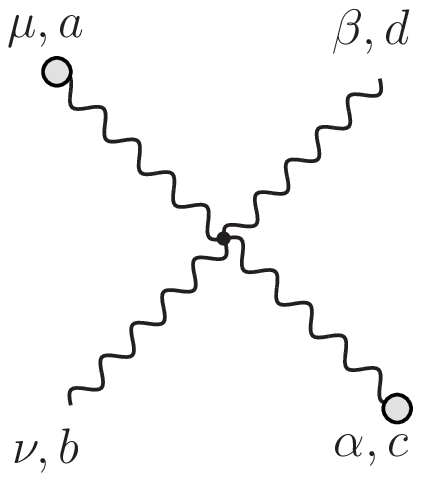}
\end{minipage}
\hspace{0.5cm}
\begin{minipage}[b]{0.75\linewidth}
\centering
 \bea
-ig^2 &&\left[f^{abx}f^{xcd}
\left(g_{\mu\alpha}g_{\nu\beta} - g_{\mu\beta}g_{\nu\alpha} +
g_{\mu\nu}g_{\alpha\beta}\right) \right. \nonumber \\
&&+\left.f^{adx}f^{xbc}
\left(g_{\mu\nu}g_{\alpha\beta} - g_{\mu\alpha}g_{\nu\beta} +
g_{\mu\beta}g_{\nu\alpha}\right)
\right. \nonumber \\
&& +\left. f^{acx}f^{xbd}
\left(g_{\mu\nu}g_{\alpha\beta} - g_{\mu\beta}g_{\nu\alpha}\right)
\right] \,.
\eea
\end{minipage}
\end{tabular}
\caption{The trilinear and quadrilinear gluon vertices in the Feynman gauge of BFM. }
\label{fig_v}
\end{figure}

After having fixed the ideas in the context of a simple Abelian model, we 
now turn to a pure Yang-Mills theory.    
In particular, we will 
study the SDE of the gluon propagator in the case of pure (quarkless) QCD,
within the PT-BFM framework.
As has been explained in detail in the recent literature~\cite{Aguilar:2006gr,Binosi:2007pi,Binosi:2008qk}, 
this latter formalism  
allows for a  gauge-invariant truncation of the SD series, 
in the sense that 
it preserves manifestly and at every step the transversality of the gluon self-energy.
\begin{figure}[!t]
\includegraphics[scale=0.75]{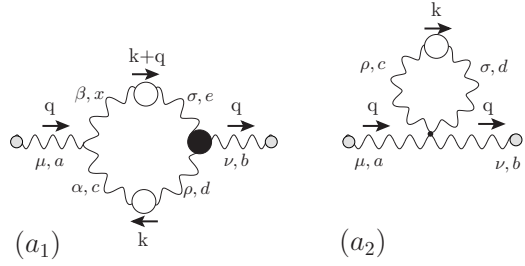}
\caption{The ``one-loop dressed'' gluonic graphs contributing to the SDE for the 
(background) gluon self-energy, $\widehat \Pi_{\mu\nu}(q)$.}
\label{Fig:groupa}
\end{figure}
%
Specifically, for the case at hand,  
we will consider only the ``one-loop dressed'' part of the gluon SDE that contains gluons 
shown in  Fig.~\ref{Fig:groupa}, 
leaving out (gauge-invariantly!) the ``one-loop dressed'' ghost contributions and 
all ``two-loop dressed'' diagrams. 
Note that the Feynman rules used to build the SD series for the  
(background) gluon self-energy, $\widehat \Pi_{\mu\nu}(q)$, are those of the 
BFM~\cite{Abbott:1980hw};  in particular, the external gluons (distinguished by the grey circles 
attached to them) are treated as if they were background gluons. 
The two tree-level vertices necessary for our analysis are given in Fig.~\ref{fig_v};
as we will see in a moment, the form of these vertices is crucial 
for obtaining from the SDE precisely the right combination of terms 
(and with the correct relative weights) that appears in  (\ref{basid}).

In order to reduce the algebraic complexity 
of the problem, we drop the longitudinal terms  
from the gluon propagators inside the integrals, i.e. we set
${\Delta}_{\alpha\beta} \to -ig_{\alpha\beta} {\Delta}$~\cite{footCLS}.
This does not compromise the transversality of $\widehat \Pi_{\mu\nu}(q)$ 
provided that we do the same on the rhs of the WI satisfied by ${\g}_{\nu\alpha\beta}$, namely we have simply
\be
q^{\nu} {\g}_{\nu\alpha\beta} = [\Delta^{-1}(k+q) -  \Delta^{-1}(k)]g_{\alpha\beta} \,,
\label{VWI}
\ee
instead of the full WI given in (\ref{VWIf}). 

Then, it is straightforward to show that the SDE corresponding to Fig.~\ref{Fig:groupa} reduces to 
\be
\widehat{\Delta}^{-1}(q)  =
q^2  + \frac{i g^2 C_{\rm {A}}}{2(d-1)} 
\bigg[ \int_k \widetilde{\Gamma}_{\mu\alpha\beta}^{(0)}
\Delta (k) \Delta(k+q) {\g}^{\mu\alpha\beta}
+  2 d^2 \int_k \!\Delta(k)\bigg],
\label{SDEm}
\ee
where $C_{\rm {A}}$ the Casimir eigenvalue 
of the adjoint representation [$C_{\rm {A}}=N$ for $SU(N)$]. 
$\widetilde{\Gamma}_{\mu\alpha\beta}^{(0)} (q,k,-k-q)$ 
is the bare three-gluon vertex in the Feynman gauge of the BFM, given in 
Fig.~\ref{fig_v}, and ${\g}_{\mu\alpha\beta}$ denotes its fully-dressed version. 

The function $\widehat{\Delta}(q)$ appearing on the lhs of (\ref{SDEm})
is the scalar part of the gluon propagator 
in the BFM, i.e. two background gluons entering; its relation to 
the self-energy $\widehat \Pi_{\mu\nu}(q)$ is the same as in (\ref{func}). 
Note that $\widehat{\Delta}(q)$   
is related to the 
standard $\Delta(q)$, defined in the $R_{\xi}$ gauges, by 
means of the powerful identity, namely $\widehat{\Delta}(q) [1+G(q^2)]^2 = \Delta(q)$, where  
$G(q^2)$ is an auxiliary two-point function~\cite{Grassi:1999tp,Binosi:2002ez}
whose dynamics have been 
studied in detail in the recent literature (see, e.g.~\cite{Aguilar:2009pp}, 
and references therein).
To further simplify the problem, without compromising its essential features, 
we will next set $G(q^2)=0$, i.e. 
we effectively assume that, inside the quantum loops, $\Delta(q)=\widehat{\Delta}(q)$.
Thus, in what follows we will be dealing with a single propagator, namely $\widehat{\Delta}(q)$, 
and will suppress the ``hats'' in order to reduce the notation.

\subsection{The three-gluon vertex}

Up until this point the analysis presented in this section 
is completely standard within the PT-BFM framework. 
At this point enters a new ingredient, namely the 
judicious Ansatz for the three-gluon vertex
which, in addition to satisfying (\ref{VWI}) will 
allow us to use the seagull identity (\ref{basid}) and get a non-vanishing and finite  $\Delta^{-1} (0)$. 
 
To begin with, let us first write $\Delta^{-1} (q)$ in the alternative form (in Minkowski space)
\be
\Delta^{-1} (q) = q^2 {H}^{-1}(q) - {\widetilde m}^2(q) \,,
\label{ombe}
\ee
The tree-level result for $\Delta^{-1} (q)$ is recovered by setting ${H}^{-1}(q)=1$ and ${\widetilde m}^2=0$. 

\begin{figure}
\bce
\includegraphics[scale=0.5]{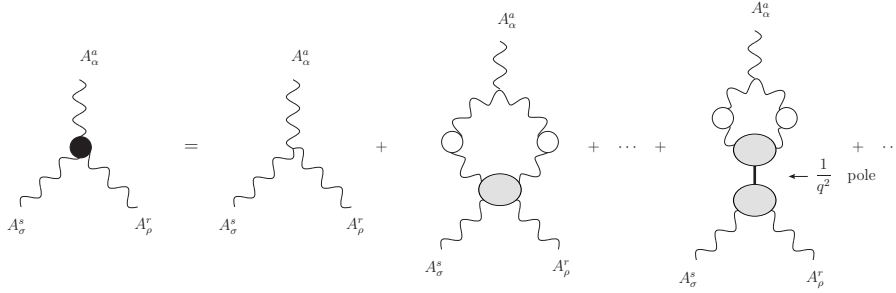}
\ece
\vspace{-0.5cm}
\caption{The SDE for the three-gluon vertex. All kernels are one-particle irreducible, and the  
$1/q^2$ pole is not kinematic but dynamical (purely non-perturbative);
physically it corresponds to a (composite) Goldstone mode, necessary for maintaining the local gauge invariance.}
\label{jsv}
\end{figure}

Then, an appropriate Ansatz for ${\g}_{\nu\alpha\beta}$ is given by    
\be
i{\g}_{\mu\alpha\beta} = 
\left[\frac{(k+q)^2 {H}^{-1}(k+q) - k^2 {H}^{-1}(k)}
{(k+q)^2-k^2}\right]\widetilde{\Gamma}_{\mu\alpha\beta}^{(0)} \,+ \,V_{\mu\alpha\beta} \,,
\label{full3}
\ee
where the term $V_{\mu\alpha\beta}$ contains the non-perturbative contributions 
due to bound-state poles associated with the Schwinger mechanism. Thus, 
$V_{\mu\alpha\beta}$ represents the term containing the $1/q^2$ pole 
on the rhs in Fig.~\ref{jsv}. 
Note that we must have   
\be
q^{\mu} V_{\mu\alpha\beta} = [{\widetilde m}^2(k) - {\widetilde m}^2(k+q)] g_{\alpha\beta}\,,
\label{mvertwi}
\ee
in order for the ${\g}_{\mu\alpha\beta}$ of Eq.~(\ref{full3}) to satisfy (by construction) the correct WI of (\ref{VWI}).

The Ansatz of (\ref{full3}) 
mimics that of Eq.~(\ref{strans_vert})  
to the extent that the first term contains the right structure 
to produce, when inserted into the first term on the rhs of (\ref{SDEm}), the  derivative term 
appearing on the lhs of (\ref{basid}). The rhs of (\ref{basid})
is already there: it is the second term on the rhs of (\ref{SDEm}), 
originating directly from the seagull diagram ($a_2$). 

Notice that the first term on the rhs of (\ref{full3}) may be expanded perturbatively, whereas 
$V_{\mu\alpha\beta}$ vanishes perturbatively to all orders. 
Qualitatively speaking, $H^{-1}(q)$ will have the form (assume for simplicity a constant ${\widetilde m}^2$)
\be
H^{-1}(q) \sim g^2 \int_0^1 dx \ln [q^2 x(x-1) + {\widetilde m^2}] + {\cal O}(g^4)\,;
\label{Hpert}
\ee
perturbatively (to all orders),  ${\widetilde m^2}=0$,  and one recovers the usual one-loop logarithm $g^2 \ln (q^2)$ 
(displaying the typical Landau pole in the IR).
Thus the role of the term $V_{\mu\alpha\beta}$ is two-fold: (i) 
it tames the Landau pole inside the dimensionless perturbative logarithm, and (ii) as can be seen directly from 
Eq.~(\ref{ombe}), it can furnish an IR-finite propagator, $\Delta^{-1} (0) = - {\widetilde m}^2(0)$ (in Minkowski space), 
provided, of course, that the equation governing ${\widetilde m^2}$ has non-trivial solutions (see next section).

An important point related to the form of ${\g}_{\mu\alpha\beta}$ is 
that the transverse (i.e. identically conserved)  component associated 
with the first term in Eq.~(\ref{full3}) has been set to zero 
[it would correspond to the term proportional to $A(k,q)$ in Eq.~(\ref{strans_vert})]. 
As is well known, such an omission,  
while of little importance in the IR, 
leads to the mishandling of overlapping divergences in the UV. This, in turn, 
spoils  the multiplicative renormalizability of the resulting SDE, which must 
be renormalized subtractively [see also discussion following Eq.~(\ref{eqhunr})].

An analogous Ansatz for the vertex $V_{\mu\alpha\beta}$ may be deduced following 
a similar philosophy. 
For example, a simple Ansatz that captures the two essential characteristics 
of having a (composite), longitudinally coupled poles, 
and satisfying the WI of  (\ref{mvertwi}) is 
\be
V_{\mu\alpha\beta} = V_{\mu\alpha\beta}^{\rm{\ell}} + V_{\mu\alpha\beta}^{\rm{t}}\,,
\label{mvert}
\ee
where
\be
V_{\mu\alpha\beta}^{\rm{\ell}} = \frac{q_{\mu}}{q^2} 
\bigg[{\widetilde m}^2(k) - {\widetilde m}^2(k+q)\bigg] g_{\alpha\beta} \,,
\label{mvert2}
\ee
and with the transverse part $V_{\mu\alpha\beta}^{\rm{t}}$ satisfying  
\be
q^{\mu}V_{\mu\alpha\beta}^{\rm{t}} =0 \,.
\ee
We emphasize that, in principle,  
the form of $V_{\mu\alpha\beta}^{\rm{t}}$ may not be chosen at will, 
but must ideally 
be determined from solving the corresponding SDE for the three-gluon 
vertex, shown schematically in Fig.\ref{jsv}.  Given 
that this task   
lies beyond our present powers, in what follows we will treat 
$V_{\mu\alpha\beta}^{\rm{t}}$ as being essentially undetermined
[see discussion before Eq.~(\ref{pim})].

We can write the vertex of (\ref{full3}) 
equivalently as 
\be
i{\g}_{\mu\alpha\beta} = 
\left[ \frac{\Delta^{-1}(k+q) -  
\Delta^{-1}(k)}{(k+q)^2-k^2} \right]\widetilde{\Gamma}_{\nu\alpha\beta}^{(0)}
\, + \,{\overline V}_{\mu\alpha\beta} \,,
\label{full4}
\ee
with 
\be
{\overline V}_{\mu\alpha\beta} = V_{\mu\alpha\beta} + V_{\mu\alpha\beta}^{\rm{r}}\,,
\ee
where 
\be
V_{\mu\alpha\beta}^{\rm{r}} = 
(2k+q)_{\mu} \left[ \frac{{\widetilde m}^2(k+q) -{\widetilde m}^2(k) }{(k+q)^2-k^2} \right]g_{\alpha\beta}\,.
\label{efvsk}
\ee
The term $V_{\mu\alpha\beta}^{\rm{r}}$ is a residual piece, acting as an additional (non-perturbative) vertex term, 
originating from forcing the vertex to assume the form of (\ref{full4}). As we will see shortly, this last way of writing ${\g}^{\nu\alpha\beta}$ 
makes the use of the basic identity of Eq.(\ref{basid}) immediate.
Thus, after these rearrangements, we have that the final non-perturbative 
effective vertex ${\overline V}_{\mu\alpha\beta}$ must be transverse,   
\be
q^{\mu} {\overline V}_{\mu\alpha\beta}  = 0 .
\ee

In summary, the vertex Ansatz proposed in (\ref{full3}) [and (\ref{full4})]
above has three important properties:
(i) satisfies identically the WIs of (\ref{VWI}), a fact that 
ensures the transversality of the resulting gluon self-energy; 
(ii) the pole term contained in $V_{\mu\alpha\beta}$
makes it possible to have a non-vanishing $\Delta^{-1} (0)$; 
(iii)  triggers the basic equation  (\ref{basid}), which, in turn, allows one to dispose of
the seagull-type terms. 
Thus, as we will see in the next subsection, the $\Delta^{-1} (0)$ obtained from the SDE is both {\it non-vanishing and finite}.   

\subsection{The implications for the SDE}

Let us now study the effect that the three-gluon vertex of (\ref{full3}) has on the SDE 
for ${\Delta}(q^2)$ given in  (\ref{SDEm}). Substituting 
for the ${\g}^{\mu\alpha\beta}$ on the rhs the expression
given in  (\ref{full4}) we obtain after simple algebra 
\be
{\Delta}^{-1}(q^2)  =
q^2  - \frac{i g^2 C_{\rm {A}}}{2(d-1)} \bigg[\Pi(q)+ \Pi_{\widetilde m}(q)\bigg] \,,
\label{hmt}
\ee
with 
\be
\Pi(q) =    
(7d-8)\, q^2 \int_k \frac{\Delta(k+q)- \Delta(k)}{(k+q)^2-k^2} 
+4d \bigg[\int_k k^2 \, \frac{\Delta(k+q)- \Delta(k)}{(k+q)^2-k^2}+ \frac{d}{2} \int_k \!\Delta(k) \bigg] \,,
\label{Piq}
\ee
and
\bea 
\Pi_{\widetilde m}(q) &=& \int_k \widetilde{\Gamma}^{(0)}_{\mu\alpha\beta}\Delta (k) \Delta(k+q)
{\overline V}^{\mu\alpha\beta}
\nonumber\\
&=& \int_k \widetilde{\Gamma}^{(0)}_{\mu\alpha\beta}\Delta (k) \Delta(k+q)
[V^{\rm{\ell}} + \{V^{\rm{t}} + V^{\rm{r}}\} ]^{\mu\alpha\beta}\,.
\label{Pim}
\eea
The term in square brackets on the rhs of (\ref{Piq}) has exactly the structure needed for employing (\ref{basid}).
In particular, using the notation introduced in (\ref{theR}), we can write  (\ref{Piq}) as   
\be
\Pi(q) = (7d-8)\, q^2 R_{\Delta}(q) + 4 d\, T_{\Delta}(q).
\label{Pi2RT}
\ee
Note the perfect balance of relative coefficients required for the precise term $T_{\Delta}(q)$ 
to emerge from the SDE.
This becomes possible within the PT-BFM framework thanks to the special vertices 
shown in Fig.~\ref{fig_v}. Instead, in the conventional SD formulation (e.g., in the $R_{\xi}$ gauges)
it would be very difficult to obtain 
the precise combination of terms needed for implementing (\ref{basid}).

Perturbatively, at one loop, $\Pi(q)$ of (\ref{Pim}) reduces to a simple and rather familiar result
[of course, $\Pi_{\widetilde m}(q)$ vanishes perturbatively, to all orders].  
Specifically, setting $\Delta(k) = 1/k^2$ on the rhs of (\ref{Piq}), it is immediate to recognize that 
the term in square brackets vanishes, since it becomes proportional to the DR 
integral $\int_k k^{-2}=0$, while the first term becomes 
\be
\Pi(q) = - (7d-8) q^2 \int_k \frac{1}{k^2 (k+q)^2} \,.
\label{Piqpert}
\ee
This is the one-loop contribution to the gluon self-energy coming from the 
graphs containing only gluons. Remember that the ghosts have been omitted without interfering 
with the transversality of the answer;
their omission 
amounts to having in front of the leading logarithm  the coefficient $(7d-8)$ instead of $(7d-6)$; 
as a result (at $d=4$)  
the coefficient of the one-loop $\beta$ function is $10 C_{\rm {A}}/ 48\pi^2$ instead of 
$11 C_{\rm {A}}/ 48\pi^2$~\cite{Abbott:1980hw,Binosi:2008qk}.


Let us now turn to the basic non-perturbative features of (\ref{hmt}).
Since by virtue of (\ref{basid}) we have that 
$T_{\Delta}(0)=0$, it is clear that $\Pi(0)=0$. Thus, the part of the calculation 
determining $\Pi(q)$ is very similar to that of scalar QED, in the sense that 
it keeps the gluon (photon) massless. 
On the other hand, the term $\Pi_{\widetilde m}(q)$, not present in the scalar QED study, 
makes it possible to have ${\Delta}^{-1}(0) \neq 0$ for the gluons. 

To see this explicitly, we focus on the $\Pi_{\widetilde m}(q)$ given in (\ref{Pim}).
The integral on the rhs of (\ref{Pim}) receives two contributions, 
one from the term containing the vertex $V^{\rm{\ell}}$ [given in (\ref{mvert2})] and 
one from the term containing the sum \{$V^{\rm{t}}+V^{\rm{r}}$\}. 
Let now us assume, for simplicity, that the (undetermined) transverse vertex $V^{\rm{t}}$ will be such that,    
when added to $V^{\rm{r}}$ [given in (\ref{efvsk})], 
will make the contribution from  \{$V^{\rm{t}}+V^{\rm{r}}$\} 
to become numerically subleading compared to that of $V^{\rm{\ell}}$. 
For instance, $V^{\rm{t}}$ could be such that the total 
contribution from  \{$V^{\rm{t}}+V^{\rm{r}}$\}
were proportional to the terms $I_2(q^2)$ and $I_4(q^2)$, shown to be subleading in (\ref{sms}). 
Then, keeping only  $V^{\rm{\ell}}$ in (\ref{Pim}), we obtain 
\bea
\Pi_{\widetilde m}(q) &=& 
-\frac{2 d }{q^2}\int_k k^2 \Delta(k)\Delta(k+q) [{\widetilde m}^2(k+q)-{\widetilde m}^2(k)]
\nonumber\\
&=& -\frac{2 d }{q^2}\int_k {\widetilde m}^2(k) \Delta(k)\Delta(k+q)[(k+q)^2-k^2]\,.
\label{pim}
\eea
Then, from the Appendix \ref{app3}, Eq.~(\ref{pimc}), we have that 
$\Pi_{\widetilde m}(0) \neq 0$, 
which, in turn, gives rise to ${\Delta}^{-1}(0) \neq 0$, as announced. 

An important consequence of this analysis is that Eq.~(\ref{hmt}) 
can be split unambiguously into two parts, one that vanishes as 
$q^2 \to 0$ and one that does not. In fact, using (\ref{ombe}) on the lhs of (\ref{hmt}), 
we can assign the two types of contributions into two separate (but coupled) equations, i.e.  
\bea
q^2 {H}^{-1}(q) &=& q^2 - \frac{i g^2 C_{\rm {A}}}{2(d-1)} \Pi(q)\,,
\label{sep1}\\
{\widetilde m}^2(q) &=& \frac{i g^2 C_{\rm {A}}}{2(d-1)} \Pi_{\widetilde m}(q)\,.
\label{sep2}
\eea
As we will see in the next section, the first equation will determine the momentum dependence of the 
effective charge, and the second the running of the gluon mass.

\section{\label{coupeq} Coupled equations for effective charge and gluon mass}

In this section we will study the system of integral equations given in Eqs.~(\ref{sep1})-(\ref{sep2}), 
under certain simplifying assumptions. 
The first step in our analysis consists in rewriting Eqs.~(\ref{sep1})-(\ref{sep2}) in terms 
of RG-invariant quantities, which will correspond to the effective charge 
and the physical gluon mass. Then, the two coupled equations will be expressed 
in terms of these two RG-invariant quantities, and will be further evaluated. 
We will assume a spectral representation for the gluon propagator [{\it viz.} Eq.~(\ref{lehmann})], 
a fact that simplifies enormously the form of the resulting equations. Finally, we will solve 
the system numerically and study the properties of the obtained solutions. 

\subsection{RG-invariant quantities}

It is well-known that, 
due to the Abelian WIs satisfied by the PT-BFM Green's functions, the 
propagator $\Delta^{-1}(q^2)$ absorbs all  
the RG logs, exactly as happens in QED with the photon self-energy.
Specifically, let us define the renormalization constants 
of the gauge-coupling  and the effective self-energy  as
\bea
g(\mu^2) &=&Z_g^{-1}(\mu^2) g_0 ,\nonumber \\
\Delta(q^2;\mu^2) & = & {Z}^{-1/2}_A(\mu^2){\Delta}_0(q^2), 
\label{conrendef}
\eea
where the ``0'' subscript indicates bare quantities.
Then, since the renormalization constants above satisfy the QED-like relation 
\be
{Z}_{g} = {Z}^{-1/2}_{A},  
\label{ptwi}
\ee
the product 
\be
{\widehat d}_0(q^2) = g^2_0 \Delta_0(q^2) = g^2 \Delta(q^2) = {\widehat d}(q^2), 
\label{ptrgi}
\ee
retains the same form before and after renormalization, i.e., it 
forms a RG-invariant \mbox{($\mu$-independent)} quantity~\cite{Cornwall:1982zr}.

For asymptotically large momenta one may extract from ${\widehat d}(q^2)$
a dimensionless quantity by writing,
\be
{\widehat d}(q^2) = \frac{\overline{g}^2(q^2)}{q^2},
\label{ddef1}
\ee
where $\overline{g}^2(q^2)$ is the RG-invariant effective charge of QCD; at one-loop
\be
\overline{g}^2(q^2) = \frac{g^2}{1+  b g^2\ln\left(q^2/\mu^2\right)}
= \frac{1}{b\ln\left(q^2/\Lambda^2_{\mathrm{QCD}}\right)}.
\label{effch}
\ee
where $\Lambda_{\mathrm{QCD}}$ denotes an RG-invariant mass scale of a few hundred ${\rm MeV}$.

The relation given in 
Eq.(\ref{ptrgi}) is true both perturbatively and non-perturbatively.
In order to realize it non-perturbatively, 
let us first set  
\be
{\widetilde m}^2(q^2) = m^2(q^2) H^{-1}(q^2) \,,
\label{rg2}
\ee
where $m^2(q^2)$ is assumed to be a RG-invariant quantity, to be identified with the dynamical gluon mass. 
Then 
\be
{\Delta}(q^2) = \frac{H(q^2)}{q^2 + m^2(q^2)}\,,
\label{rg3}
\ee
and from the requirement that  
$g^2 {\Delta}(q^2)$ must be RG-invariant we have that 
\be
g^2 H(q^2) = {\overline g}^2(q^2)\,.
\label{rg1}
\ee
Therefore, we finally arrive at the RG-invariant combination 
\be
{\widehat d}(q^2)\equiv g^2{\Delta}(q^2) = {\overline g}^2(q^2)\bar\Delta(q^2)\,,
\label{rg4}
\ee
with
\be
\bar\Delta(q^2)= \frac{1}{q^2 + m^2(q^2)}\,.
\label{rg5}
\ee
Evidently the dimensionful RG-invariant quantity ${\widehat d}(q^2)$ is decomposed 
into the product of two individually RG-invariant quantities, the dimensionful part $[q^2 + m^2(q^2)]^{-1}$
corresponding to a massive propagator (with a running mass), and the dimensionless 
${\overline g}^2(q^2)$ corresponding to the running coupling (effective charge).  

\subsection{The equation for the effective charge}

Even though in principle the analysis may be carried out 
using systematically the formulas of Appendix (\ref{app1})  
without imposing any additional constraints on $\Delta$, 
the presence of the derivatives makes the numerical treatment rather cumbersome.
Instead, as shown in Appendix (\ref{app2}), the use of the spectral representation for $\Delta$
results in a spectacular simplification. 
  
Specifically, assuming that $\Delta$ can be written as in (\ref{lehmann}), and using the expressions in (\ref{sf4}), 
we have that 
\be
-i\Pi(q) = 
\frac{(7d-8)}{16\pi^2} q^2 \left[\int^{q^2/4}_{0}\!\!\!dz  \left(1-\frac{4z}{q^2}\right)^{\!\! 1/2}\!\!\!\! \Delta(z) 
- {\cal C}\right]
+ \frac{4d}{16\pi^2} \int^{q^2/4}_{0}\!\!\!dz \,z \,\left(1-\frac{4z}{q^2}\right)^{\!\!1/2}\!\!\!\! \Delta(z) \,,
\label{Pi2RT2}
\ee

The equation for the effective charge, ${\alpha}(q^2) = {\overline g}^2(q^2)/4\pi$, will be derived from 
(\ref{sep1}) after substitution of (\ref{Pi2RT2}).  
At this point we go to Euclidean momenta; specifically 
we set  $q^2 = -q^2_{\chic E}$, with  $q^2_{\chic E} >0$ the positive square of a 
Euclidean four-vector, and define the Euclidean propagator as 
$\Delta_{\chic E} (q^2_{\chic E}) = -\Delta (-q^2_{\chic E})$ 
(we suppress the subscript ``E'' in what follows).  
Then, from (\ref{sep1}) we have 
\be
{H}^{-1} (q^2) = K +  \tilde{b} g^2
\left[\int^{q^2\!/\!4}_{0}\!\!\!dz  \left(1+ \frac{4z}{5q^2}\right) \left(1-\frac{4z}{q^2}\right)^{\!\! 1/2}\!\!\!\! \Delta(z) 
- {\cal C}\right]\,, 
\label{eqhunr}
\ee
where $\tilde{b} = 10 C_{\rm {A}}/ 48\pi^2 $; the discrepancy from the correct 
factor $b = 11 C_{\rm {A}}/ 48\pi^2 $, namely the first coefficient 
of the QCD one-loop $\beta$-function, is due to the (gauge-invariant!) omission 
of the ghost loops.    
The (infinite) constant $K$ is the gluon wave-function renormalization, introduced in order 
to make the equation finite, i.e. eliminate the infinite constant ${\cal C}$.  
Note that, as is typical in this type of SDE analysis, 
the renormalization is carried out subtractively instead of multiplicatively.
This is ultimately connected with the fact that the transverse part of the three-gluon vertex 
is undetermined by the gauge technique (see discussion in the previous section). 

The constant  $K$  may be determined from (\ref{eqhunr}) by imposing 
a renormalization condition on the function ${H}^{-1} (q^2)$. Specifically, using the MOM-type
of condition ${H}^{-1} (\mu^2) =1 $, we have that $K$ is given by 

\be
K= 1- \tilde{b} g^2
\left[\int^{\mu^2\!/\!4}_{0}\!\!\!dz  \left(1+ \frac{4z}{5\mu^2}\right) 
\left(1-\frac{4z}{\mu^2}\right)^{\!\! 1/2}\!\!\!\! \Delta(z) 
- {\cal C}\right]\,.
\label{renc}
\ee

Inserting the expression for $K$ given in  (\ref{renc}) back into (\ref{eqhunr}), we obtain 
the renormalized equation  

\be 
{H}^{-1} (q^2) =  1 +  \tilde{b} g^2
\left[\int^{q^2\!/\!4}_{0}\!\!\!dz  \left(1+ \frac{4z}{5q^2}\right) \left(1-\frac{4z}{q^2}\right)^{\!\! 1/2}\!\!\!\! \Delta(z)
 - \int^{\mu^2\!/\!4}_{0}\!\!\!dz  \left(1+ \frac{4z}{5\mu^2}\right) 
\left(1-\frac{4z}{\mu^2}\right)^{\!\! 1/2}\!\!\!\! \Delta(z)\right] \,,
\label{eqhr}
\ee

In order to derive the equation for the effective charge ${\alpha}(q^2) = {\overline g}^2(q^2)/4\pi$,  
use the relation between $H(q^2)$ and ${\overline g}^2(q^2)$ 
given in (\ref{rg1}),  to cast (\ref{eqhr}) in the form 
\be
\frac{1}{{\overline g}^2(q^2)} =  \frac{1}{{\overline g}^2(\mu^2)} +  \tilde{b}
\left[\int^{q^2\!/\!4}_{0}\!\!\!dz \left(1+ \frac{4z}{5q^2}\right) \left(1-\frac{4z}{q^2}\right)^{\!\! 1/2}\!\!\!\! \Delta(z)  
- \int^{\mu^2\!/\!4}_{0}\!\!\!dz \left(1+ \frac{4z}{5\mu^2}\right) \left(1-\frac{4z}{\mu^2}\right)^{\!\! 1/2}\!\!\!\! \Delta(z)\right]\,.
\label{eqalpha}
\ee
Note that again, because of the mishandling of the transverse part of the three-gluon vertex, 
the rhs of (\ref{eqalpha}) is not RG-invariant. The simplest way to remedy this (by hand) 
is to replace $\Delta(z) \to \bar\Delta(z)$ of Eq.(\ref{rg5}).
Thus, we arrive at 
\be
\frac{1}{{\overline g}^2(q^2)} =  \frac{1}{{\overline g}^2(\mu^2)} 
+ \tilde{b}\left[\int^{q^2\!/\!4}_{0}\!\!\!dz \left(1+ \frac{4z}{5q^2}\right) \left(1-\frac{4z}{q^2}\right)^{\!\! 1/2}\!\!\!\! \bar\Delta(z)  
- \int^{\mu^2\!/\!4}_{0}\!\!\!dz  \left(1+ \frac{4z}{5\mu^2}\right)\left(1-\frac{4z}{\mu^2}\right)^{\!\! 1/2}\!\!\!\! \bar\Delta(z)\right]\,.
\label{eqalphargi}
\ee
Note that
\be
\frac{1}{{\overline g}^2(0)} =  \frac{1}{{\overline g}^2(\mu^2)} 
- \tilde{b} \left[  
\int^{\mu^2\!/\!4}_{0}\!\!\!dz  \left(1+ \frac{4z}{5\mu^2}\right)\left(1-\frac{4z}{\mu^2}\right)^{\!\! 1/2}\!\!\!\! \bar\Delta(z)
\right]\,.
\label{eqalpha0}
\ee

\subsection{The equation for the gluon mass}

Let us now turn to the dynamical equation governing the evolution of the mass.
From (\ref{sep2}) we obtain 
\be
{\widetilde m}^2 (q^2)= \frac{2}{5}\tilde{b} g^2  [I_1(q^2)+I_2(q^2)+I_3(q^2)+I_4(q^2)]\,, 
\label{eqm}
\ee
where the terms $I_i(q)$ are  given in (\ref{sms}).
According to the discussion in (\ref{app3}), the terms $I_2(q^2)$ and $I_4(q^2)$ 
are subleading both the  IR and the UV, and may be therefore safely neglected 
to a first approximation. Then, keeping only  $I_1(q^2)$ and $I_3(q^2)$, we have
\be
{\widetilde m}^2 (q^2)= \frac{2}{5}\tilde{b} g^2
\bigg[
\Delta(q^2) \int_0^{q^2} dy y {\widetilde m}^2(y) \Delta(y) 
- \frac{1}{2} \,\int_{q^2}^{\infty}  dy y^2 \Delta^2(y) [{\widetilde m}^{2}(y)]^{\prime}\bigg]\,.
\ee
The next step is to rewrite this equation in terms of the RG-invariant quantities. 
Using (\ref{rg2})-(\ref{rg1}), we have that 
\be
\frac{m^2(q^2)}{{\overline g}^2(q^2)} = \frac{2}{5}\tilde{b} g^2 
\bigg[
\Delta(q^2) \int_0^{q^2}\!\! dy y \Delta(y) [m^2(y)/{\overline g}^2(y)] 
- \frac{1}{2} 
\,\int_{q^2}^{\infty}  \!\! dy y^2 \Delta^2(y) [m^2(y)/{\overline g}^2(y)]^{\prime}
\bigg]\,.
\ee
Given that ${\overline g}^2(y)$ is expected to be a much slower varying function of 
the momentum compared to $m^{2}(y)$, both in the UV and the IR, we will simplify the analysis 
by neglecting the derivative $[{\overline g}^2(y)]^{\prime}$ next to $[m^{2}(y)]^{\prime}$.
Then, we have that 
\be
\frac{m^2(q^2)}{{\overline g}^2(q^2)} = \frac{2 \tilde{b}}{5}
\bigg[
\Delta(q^2) \int_0^{q^2} \!\!dy y m^2(y)\{g^2\Delta(y)/{\overline g}^2(y)\}
- \frac{1}{2} 
\int_{q^2}^{\infty} \!\! dy y^2 [m^{2}(y)]^{\prime} \Delta(y)\{g^2\Delta(y)/{\overline g}^2(y)\}
\bigg]\,,
\label{ppg0}
\ee
which, after using (\ref{rg4}),  becomes 
\be
\frac{m^2(q^2)}{{\overline g}^2(q^2)} = \frac{2 \tilde{b}}{5}
\bigg[
\Delta(q^2) \int_0^{q^2} dy y m^2(y) \bar\Delta(y)
- \frac{1}{2} 
\int_{q^2}^{\infty}  dy y^2 [m^{2}(y)]^{\prime}\Delta(y)\bar\Delta(y) 
\bigg]\,.
\label{ppg}
\ee
Finally, the rhs of (\ref{ppg}) is made RG-invariant by setting  
$\Delta(q^2)\to \bar\Delta(q^2)$ and $\Delta(y)\to {\widehat d}(y)$, thus obtaining   
\be
\frac{m^2(q^2)}{{\overline g}^2(q^2)} = \frac{2 \tilde{b}}{5} 
\bigg[
\bar\Delta(q^2) \int_0^{q^2} \!\!\! dy y m^2(y) \bar\Delta(y) 
\,-\, \frac{1}{2} 
\int_{q^2}^{\infty}\!\!\! dy y^2 {\bar\Delta}^2(y) {\overline g}^2(y) [m^{2}(y)]^{\prime} 
\bigg]\,.
\label{ppgtm}
\ee
Let us now study the behavior of the solutions of (\ref{ppgtm}) for asymptotically large $q^2$; 
in this limit we set ${\bar\Delta}(x)\to 1/x$ and ${\bar\Delta}(y)\to 1/y$. 
Then, the equation reduces to   
\be
m^2(q^2) \ln q^2 =  \frac{2}{5} 
\bigg[ \frac{1}{q^2} \int_0^{q^2} dy \, m^2(y) \, - \,\frac{1}{2} 
\int_{q^2}^{\infty} \!\!\! dy {\overline g}^2(y) [m^{2}(y)]^{\prime}
\bigg]\,.
\label{pls}
\ee
It is relatively straightforward to establish that the asymptotic solutions of (\ref{pls}) display power-law running.
Indeed, substituting on both sides of (\ref{pls}) a $m^2(q^2)$ of the form  
\be
m^2(q^2)  = \frac{\lambda_0^4}{q^2} (\ln q^2)^{\gamma-1}\,, 
\label{plr}
\ee
it is easy to recognize 
that the second term on the rhs of (\ref{plr}) is subleading. Indeed, 
in the absence of ${\overline g}^2(y) = (b\ln y)^{-1}$ the integrand is a total derivative, 
which yields to the rhs simply a term $\frac{1}{2} m^{2}(q^2)$; this is suppressed, because it is not multiplied by a  
$\ln q^2$. The presence of ${\overline g}^2(y)$ suppresses this integral even further.
Specifically, integration by parts and use of the second equation in (\ref{intm23}) yields 
\be 
\int_{q^2}^{\infty} \!\!\! dy {\overline g}^2(y) [m^{2}(y)]^{\prime}
= - m^2(x) {\overline g}^2(x) + {\cal O}\left( 1/\ln x\right)\,,   
\ee
which is indeed further suppressed by an extra logarithm.   

Thus, using the elementary integral 
\be
\int  \frac{dy}{y\, (\ln y)^{1+a}} = - \frac{1}{a \, (\ln y)^{a}}\,,
\label{elint}
\ee
[first equation in (\ref{intm23})], 
we have that (\ref{plr}) is a solution of (\ref{pls}) provided that 
\be
\gamma = \frac{2}{5} \,,
\ee
and so, the asymptotic solution has power-law running, given by 
\be
m^2(q^2)  = \frac{\lambda_0^4}{q^2} (\ln q^2)^{-3/5}\,.
\label{plras}
\ee

Finally, if we were to assume the approximate validity of (\ref{plras}) for the entire range of momenta, 
we can set $[m^{2}(y)]^{\prime} \approx - m^{2}(y)/y$; that way,  
we convert (\ref{pls}) from an integro-differential equation to 
the simpler integral equation 
\be
\frac{m^2(q^2)}{{\overline g}^2(q^2)} = \frac{2 \tilde{b}}{5} 
\bigg[
\bar\Delta(q^2) \int_0^{q^2} \!\!\! dy y m^2(y) \bar\Delta(y) 
\,+\, \frac{1}{2} 
\int_{q^2}^{\infty}\!\!\! dy y {\bar\Delta}^2(y) {\overline g}^2(y) m^{2}(y) 
\bigg]\,.
\label{ppgtmfin}
\ee

\begin{figure}[!t]
\begin{center}
\includegraphics[scale=0.9]{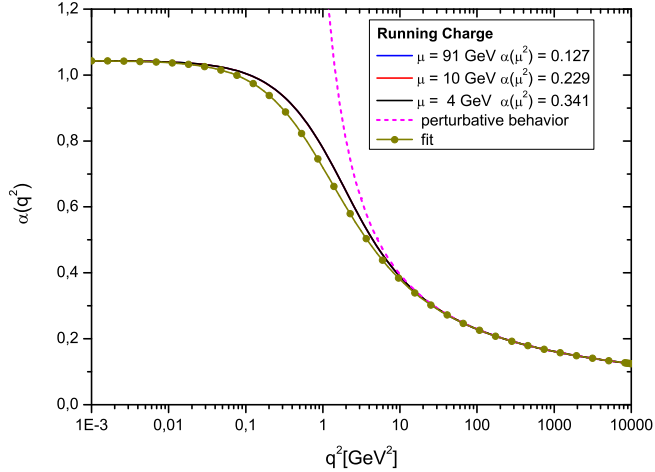}
\end{center}
\vspace{-1.0cm}
\caption{Numerical solutions for the effective charge obtained from Eq.~(\ref{eqalphargi}), renormalized at three different points:  $\mu = 4 \,\mbox{GeV}$ and \mbox{$\alpha(\mu^2)=0.341$} (black curve), $\mu = 10 \,\mbox{GeV}$ and \mbox{$\alpha(\mu^2)=0.229$} (red curve), $\mu = 91 \,\mbox{GeV}$ and \mbox{$\alpha(\mu^2)=0.127$} (blue curve). The three curves practically coincide, 
showing that indeed $\alpha(q^2)$ is independent of the renormalization point chosen.
The dashed curve (magenta) is the perturbative one-loop behavior, and the brown line
with circles depicts the fit of Eq.(\ref{fit_coup}), for $t=3.7$.}
\label{plot1}
\end{figure}

\subsection{Solving the system numerically}

We will next discuss the numerical solutions for 
the system of integral equations, namely   
(\ref{eqalphargi}) and (\ref{ppgtmfin}) coupled together.

We solve numerically the two coupled integral equations, 
renormalizing them at three different
points, namely $\mu=\{4,10,91\}\,\mbox{GeV}$, with $\alpha(\mu^2)= g^2(\mu)/4\pi =\{0.341,0.229,0.127\}$, respectively. 
In Fig.~\ref{plot1}, we show the results for $\alpha(q^2)$;  there we see clearly that 
the three curves merge practically into a single one, thus 
confirming numerically the $\mu$-independence of $\alpha(q^2)$, expected on formal grounds. 
These three curves may be  accurately fitted  by the 
physically motivated functional form~\cite{Cornwall:1982zr}, namely   
\be
\alpha(q^2) = \frac{1}{4\pi\tilde{b}\ln[(q^2+tm_0^2)/\Lambda^2]} \,,
\label{fit_coup}
\ee
with $t=3.7$ and $\Lambda= 645\,\mbox{MeV}$ [see caption of Fig.~\ref{plot1}]. 
%
\begin{figure}[!t]
\begin{center}
\includegraphics[scale=0.9]{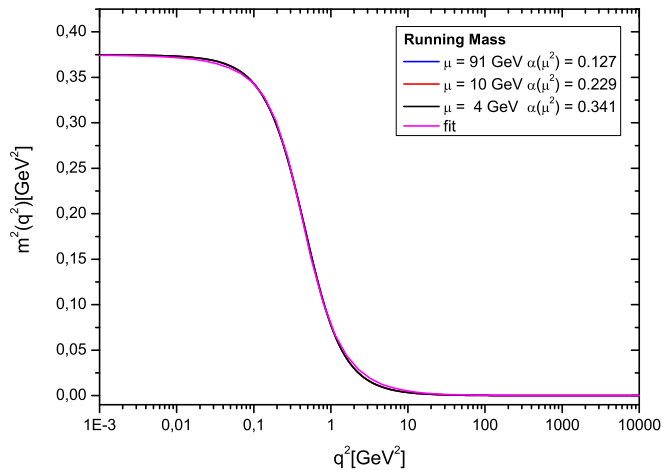}
\end{center}
\vspace{-1.4cm}
\caption{ The effective gluon mass, $m^2(q^2)$, for the same renormalization points used in  Fig.~(\ref{plot1}). 
Evidently, the three curves merge into a single one, showing that the numerical solutions 
are independent of the renormalization point. 
The continuous line in magenta is the fit of Eq.~(\ref{fit_mass}) with $\rho_1=-1/2$ and $\rho_2=5/2$.}
\label{plot2}
\end{figure}

 In Fig.~\ref{plot2}, we show the dynamical gluon mass,  $m^2(q^2)$, 
obtained as solution of  Eq.~(\ref{ppgtmfin}) at the same renormalization points of Fig.~\ref{plot1}.  
Once again, this figure shows us that $m^2(q^2)$ is also a RG-invariant quantity, since
the three curves, obtained using the three different (and quite disparate) renormalization points, are practically on top 
of each other. The behavior of $m^2(q^2)$ in the entire range of momenta can be accurately described by the following parametrization
\be
m^2(q^2)=\frac{m_0^4}{q^2+m_0^2}\left[\ln\left(\frac{q^2+f(q^2,m_0^2)}{\Lambda^2}\right)\bigg/
\ln\left(\frac{f(0,m_0^2)}{\Lambda^2}\right)\right]^{-3/5} \,,
\label{fit_mass}
\ee
where the function 
\be
f(q^2,m_0^2)= \rho_1 m_0^2 + \rho_2 \frac{m_0^4}{q^2 + m_0^2}\,,
\label{fmass}
\ee
with $\rho_1=-1/2$, $\rho_2=5/2$, and $m_0=612\,\mbox{MeV}$.
Notice that in the UV asymptotic limit the above expression goes over to that of Eq.(\ref{plras}), as it should. 
\begin{figure}[!b]
\begin{center}
\includegraphics[scale=0.9]{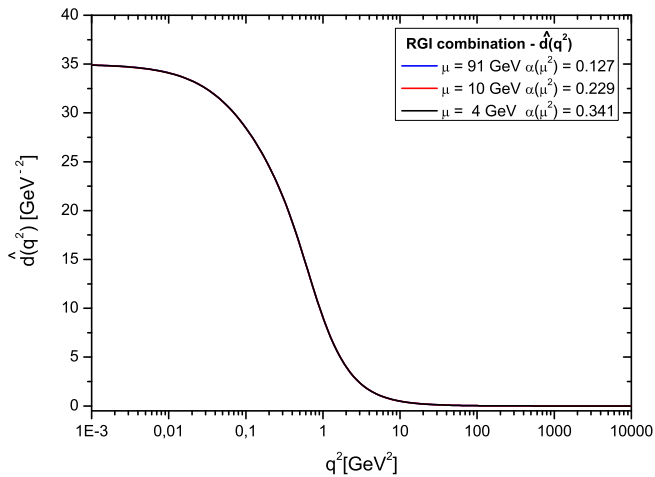}
\end{center}
\vspace{-1.2cm}
\caption{The renormalization-group invariant product $\widehat{d}(q^2)$ obtained by combining the 
results for $\alpha(q^2)$ and $m^2(q^2)$ according to Eq.~(\ref{rg4}).}
\label{plot3}
\end{figure}

Finally,  we turn to the RG-invariant quantity $\widehat{d}(q^2)$,  which appears in a natural way in all physical process involving gluon exchange.
With the help of Eq.~(\ref{rg4}) we can construct $\widehat{d}(q^2)$ out of the numerical solutions for $\alpha(q^2)$ and $m^2(q^2)$; the result is 
shown in Fig.~{\ref{plot3}}. Obviously, since $\widehat{d}(q^2)$ is built out of two quantities that are individually independent of $\mu$, 
it too turns out to be  $\mu$-independent; this property is clearly observed in  Fig.~\ref{plot3}.   

\section{\label{concl} Discussion and Conclusions}

In this article we have demonstrated how to obtain a finite gluon mass 
from the SDE of QCD, formulated in the PT-BFM framework. 
Obtaining a finite mass without the appearance of seagull divergences has been a long-standing problem, that has afflicted 
all related studies for a number of years. 
The key observation that leads to the solution of this problem 
is that a judicious Ansatz for the three-gluon vertex
eliminates all seagull divergences by means of a basic identity, 
valid in dimensional regularization.  
In retrospect one realizes that 
the problem of seagull divergences is not intrinsic to this approach, 
but has rather been caused by the inadvertent 
mismatch of two field theoretic mechanisms, 
induced by an imperfect Ansatz for the vertex. Specifically, the Schwinger 
mechanism, which requires the appearance of massless poles in the three-gluon vertex, 
distorts the mechanism responsible for the 
cancellation of the seagull divergences, {\it unless} the poles enter into the gluon vertex in a very particular way.

The procedure described in the present work furnishes eventually two separate but coupled 
equations for the QCD effective charge and the gluon mass, 
which, when solved simultaneously, yield a unique answer for both quantities.
This is a considerable improvement over the existing approaches (e.g.~\cite{Cornwall:1982zr,Aguilar:2006gr,Aguilar:2007ie}) where 
one had only one dynamical equation, determining $\widehat{d}(q)$, which was subsequently 
decomposed according to Eq.~(\ref{rg4}), in order to obtain (not without a certain ambiguity) 
the effective charge and gluon mass.    
It should be emphasized, however, that even though the 
values obtained for $\alpha(0)$ and $m_0$ are very reasonable, they are not directly comparable  
with the values obtained from phenomenological studies, due to the fact that the gauge used in this work
(``stagnant'' or ``generalized Feynman'') is {\it not} the canonical Feynman gauge of the BFM, which, as is well-known, 
furnishes the PT effective charge and gluon mass. This may account, in retrospect, for the 
slightly elevated value of $\alpha(0)\approx 1$ obtained here, compared to a value of $0.5-0.7$, 
found in recent theoretical analysis~\cite{Cornwall:2009ud}, 
and various phenomenological studies~\cite{Aguilar:2002tc,Aguilar:2001zy}

The fully dressed three-gluon vertex ${\g}_{\mu\alpha\beta}$
used in Eq.~(\ref{full3}) satisfies (by construction) the WI given in 
Eq.~(\ref{VWI}), or, after restoring the longitudinal terms, the full WI of Eq.~(\ref{VWIf}), 
which is crucial for ensuring the transversality of the gluon self-energy.  
However, this Ansatz used for  ${\g}_{\mu\alpha\beta}$ 
is still incomplete from the point of view
of full Bose symmetry, in the sense that it does not satisfy the correct STI when contracted with respect 
to the two other legs; we remind the reader that the other two legs (internal lines, irrigated by virtual momenta) 
correspond to  ``quantum'' gluons, 
as opposed to the external (background) gluon, where the physical momentum $q$ enters.  
The corresponding STI satisfied when contracting ${\g}_{\mu\alpha\beta}$ with respect to the quantum legs 
is a variant of the well-known 
STI satisfied by the conventional vertex~\cite{Ball:1980ax}, and has been derived in~\cite{Binosi:2008qk}.
The construction of a three-gluon vertex satisfying the correct WI and STIs is currently under investigation, 
and we hope to report the results in the near future.

An additional technical issue related to the form of ${\g}_{\mu\alpha\beta}$ is 
the omission of the  identically conserved part, 
as already mentioned in the corresponding sections. It would be most interesting to 
extend the QED construction of~\cite{King:1982mk,Kizilersu:2009kg}
to the case of the three-gluon vertex ${\g}_{\mu\alpha\beta}$. Such a task, however, 
appears to be of formidable logistic complexity,  
given that there are thirteen linearly independent tensorial structures, with their corresponding  
form-factors. The techniques and the special tensorial bases introduced in~\cite{Binger:2006sj} 
may prove useful for simplifying such a task.

In a similar spirit, the part $V^{\alpha\beta\gamma}$ of the vertex containing the massless poles, thus triggering the 
Schwinger mechanism, should also be appropriately extended, to satisfy the 
correct WI with respect to all three legs. A prime candidate for this role is 
the vertex proposed in~\cite{Cornwall:1985bg}, given by 
\be
V^{\alpha\beta\gamma}(k_1,k_2,k_3) = \frac{k_1^{\alpha} k_2^{\beta}(k_1-k_2)^{\mu}}{2k_1^{2} k_2^{2}}
P_{\mu}^{\gamma}(k_3) m^2(k_3) - \frac{k_3^{\gamma}}{k_3^{2}} 
\left[m^2(k_2)-m^2(k_1)\right]P^{\alpha}_{\mu}(k_1)P^{\mu\beta}(k_2) + {\rm cp}\,, 
\label{meta}
\ee
where ``${\rm cp}$'' denotes ``cyclic permutations''.  Note that a vertex such as (\ref{meta})
will furnish 
a concrete (but still not unique) expression for $V^{\rm{t}}$, which, in turn, will allow one to 
scrutinize some of the assumptions made in Section~\ref{fgmg}.  
Of course, for self-consistency, one should perform the 
analysis in, e.g., the Landau or conventional Feynman gauges, rather then the 
``generalized'' Feynman gauge employed here; calculations in this direction are already in progress.

\appendix

\section{\label{app1} Some useful relations}

Let us define the following quantities, 
\bea
R_f(q) & \equiv  & \int_k  \frac{f(k+q)- f(k)}{(k+q)^2-k^2} \,,
\nonumber\\
T_f(q) & \equiv  & \int_k k^2 \, \frac{f(k+q)- f(k)}{(k+q)^2-k^2} + \frac{d}{2} \int_k \! f(k) \,,
\label{theR}
\eea
for an arbitrary function $f(x)$ that is finite at the origin. 

Let us define $q^2 =x$, $k^2 =y$, $(k+q)^2 =z$, 
and 
let us write the (Euclidean) integration measure [$d^d k = i d^d k_{\chic E}$] in spherical 
coordinates
\be
\int d^{d}k_{\chic E}= 2\pi \!\!\int_{0}^{\pi} \!\!\! d\theta (\sin\theta)^{d-2} \, 
\int_{0}^{\infty}\!\!\! dy\ y.
\label{spher}
\ee
We then have that 
$z= y+x+2 \sqrt{xy} \cos\theta $, and we define    
$ w \equiv z-y= x+2 \sqrt{xy} \cos\theta $. Finally,   
recall the elementary integral 
\be
\int_{0}^{\pi} \!\!\! d\theta\sin^m\theta \cos^n\theta =
\left\{
\begin{array}{ll}
\frac{\Gamma\left(\frac{m+1}{2}\right) 
\Gamma\left( \frac{n+1}{2}\right) }{\Gamma\left( \frac{m+n+2}{2}\right) } \,, & n= 2k \\
0 \,, & n= 2k+1
\end{array}
\right.
\label{intsin}
\ee
$R_f(q)$ and $T_f(q)$ may be expanded systematically as a power series in $q^2$. 
To that end we consider the Taylor expansion of $f(z)$ around $w=0$,  
which gives (we are assuming finite derivatives at the origin),  
\be 
\frac{f(z)- f(y)}{w} = f^{\prime}(y) + 
\frac{w}{2!}   f^{\prime\prime}(y) + \frac{w^2}{3!}  f^{\prime\prime\prime}(y) + ...
\label{tayl}
\ee
where the primes denote differentiations with respect to $y$. 
Then, one must collect pieces of a given order in $q^2$ from the various powers of $w$, 
using (\ref{intsin}).  

It is clear, for example, that 
when the term  $f^{\prime}(y)$ 
on the rhs of (\ref{tayl}) is inserted into  $T_f(q)$ 
generates the seagull identity (\ref{basid}), while all remaining terms proportional to positive powers of $w$;
so,  
\be
T_f(0)=0.
\label{tf0}
\ee
As a second example, we determine the term of $R_f(q)$  
linear in $q^2$, to be denoted by $R_f^{(1)}(q)$; to accomplish this 
one must collect the appropriate contributions coming from 
both the second and the third term on the rhs of (\ref{tayl}). Using (\ref{basid}) one then obtains
\be
R_f^{(1)}(q) = q^2 \int_k \left[\frac{1}{2} f^{\prime\prime}(k^2) + \frac{1}{6} k^2 f^{\prime\prime\prime}(k^2)\right] \,,
\label{rf1}
\ee
or after partial integration, assuming that $[yf^{\prime}]^{\infty}_0 =0$ and $[y^2 f^{\prime\prime}]^{\infty}_0 =0$ 
(valid when $f(0)$ is finite, and $f(y) \sim y$ (or faster) at infinity), we have 
\be
R_f^{(1)}(q) = \frac{c}{6} q^2 f(0) \,,
\ee
\label{rf1b}
where $c \equiv i/16\pi^2$ .

\section{\label{app2} The spectral representation}

Let us consider a simple massive tree-level propagator, 
\be
d_m(q) = \frac{1}{q^2 -m^2}\,,
\ee
and set $f=d_m$ directly into 
(\ref{theR}). Turns out that both $R_{d_m}(q)$ and 
$T_{d_m}(q)$ can be calculated exactly; 
specifically, using that 
\be
\frac{{d}_m(k+q) - {d}_m(k)}{(k+q)^2-k^2}= 
- {d}_m(k) {d}_m(k+q) \,,
\label{ff1}
\ee
it is elementary to show that 
\be
R_{d_m}(q) = 
c\left[\int_{0}^{1} dx \ln\left(1 + \frac{q^2 x(x-1)}{m^2} \right)\, - \, C_m \right] \,,
\label{sf1}
\ee
where the cutoff-dependent constant $C_m$ is given in dimensional regularization by  
\be
C_m= \frac{2}{\epsilon} - \gamma - \ln\left(\frac{m^2}{4\pi\mu^2}\right)\,.
\ee
Evidently, $R_{d_m}(0) = - c C_m$. 
As an additional check, note that 
the term of $R_{d_m}(q)$ linear in $q^2$, obtained by Taylor-expanding (\ref{sf1}), is given by  
\be 
R_{d_m}^{(1)}(q) = - \frac{c}{6}  \frac{q^2}{m^2}\,,
\ee
which coincides with the result obtained when 
substituting $f(0) = d_m(0)=-\frac{1}{m^2}$  in the general formula of (\ref{rf1b}).

For $T_{d_m}(q)$ we have, using the identities (\ref{id1})-(\ref{id2}), 
\bea
T_{d_m}(q) &=& m^2  \int_k \frac{q^2 + 2 q\cdot k}{(k^2-m^2)^2 [(k+q)^2-m^2]}
\nonumber\\
&=&  -c m^2 q^2 \int_0^1 dx \frac{x(2x-1)}{q^2 x (1-x) + m^2}
\nonumber\\
&=& c m^2  
\int_{0}^{1} dx \ln\left(1 + \frac{q^2 x(x-1)}{m^2} \right)\,.
\label{sf2}
\eea
Evidently,  $T_{d_m}(0)=0$, in agreement with (\ref{tf0}).

The results of (\ref{sf1})-(\ref{sf2}) may be used in a more general way. 
Specifically, if we assume a spectral representation for the gluon propagator~\cite{Cornwall:1982zr,posit}, namely 
\begin{equation}
\Delta (q^2) = \int \!\! d \lambda^2 \, \frac{\rho\, (\lambda^2)}{q^2 - \lambda^2 + i\epsilon}\,;
\label{lehmann}
\end{equation}
then from (\ref{sf1}) (with $m \to \lambda$) we have 
\bea
R_{\Delta}(q) &=&  c
\left[\int \!\! d \lambda^2 \, \rho\, (\lambda^2) \int_{0}^{1} dx \ln\left(1 + \frac{q^2 x(x-1)}{\lambda^2} \right)
-{\cal C}\right] \,,
\nonumber\\
T_{\Delta}(q) &=&  c \int \!\! d \lambda^2 \, \rho\, (\lambda^2) \lambda^2 
\int_{0}^{1} dx \ln\left(1 + \frac{q^2 x(x-1)}{\lambda^2} \right)\,,
\label{sf3}
\eea
where 
\be
{\cal C} = \int \!\! d \lambda^2 \, \rho\, (\lambda^2) C_{\lambda}\,.
\ee
Note that after the renormalization 
of the corresponding SDE (see Section~\ref{coupeq}) ${\cal C}$ will eventually drop out.
Then, the use of the following identities~\cite{f5}
\bea
\int \!\! d \lambda^2 \, \rho\, (\lambda^2) 
\int_{0}^{1} dx \ln\left(1 + \frac{q^2 x(x-1)}{\lambda^2} \right)
&=& \int^{q^2\!/\!4}_{0}\!\!\!dz  \left(1-\frac{4z}{q^2}\right)^{\!\!1/2} \!\!\!\Delta(z)\,,
\nonumber\\
\int \!\! d \lambda^2 \, \rho\, (\lambda^2) \,\lambda^2
\int_{0}^{1} dx\, \ln\left(1 + \frac{q^2 x(x-1)}{\lambda^2} \right)
&=& 
\int^{q^2\!/\!4}_{0}\!\!\!dz \,z \,\left(1-\frac{4z}{q^2}\right)^{\!\!1/2} \!\!\!\Delta(z)\,,
\label{Integrals}
\eea
allows one to cast $R_{\Delta}(q)$ and $T_{\Delta}(q)$ 
again as an integral containing the gluon propagator $\Delta$, namely 
 \bea
R_{\Delta}(q) &=&  c
\left[\int^{q^2\!/\!4}_{0}\!\!\!dz  \left(1-\frac{4z}{q^2}\right)^{\!\!1/2} \!\!\!\Delta(z)
-\,{\cal C}\right]\,,
\nonumber\\
T_{\Delta}(q) &=&  c \int^{q^2\!/\!4}_{0}\!\!\!dz \,z \,\left(1-\frac{4z}{q^2}\right)^{\!\!1/2} \!\!\!\Delta(z)\,.
\label{sf4}
\eea
Note that the simple change of variables $t=4z/q^2$ allows one to cast $R_{\Delta}(q)$ and $T_{\Delta}(q)$
in the alternative form 
 \bea
R_{\Delta}(q) &=&  c \left[(q^2/4) \int^{1}_{0} dt \, (1-t)^{1/2} \Delta(t q^2/4) -\,{\cal C}\right]\,,
\nonumber\\
T_{\Delta}(q) &=&  c \,(q^2/4)^{2}  \int^{1}_{0} dt\, t\, (1-t)^{1/2} \Delta(t q^2/4)\,,
\label{sf41}
\eea
which makes the identification of the IR behavior of these quantities immediate, and is 
particularly useful for their numerical treatment.

\section{\label{app3}The mass equation}

Let us consider (in Euclidean space)
the integral appearing on the rhs of (\ref{pim}), to be denoted by $I(q)$. 
We have 
\be
I(q) = \frac{1}{q^2}\int_{k_{\chic E}} {\widetilde m}^2(k) \Delta(k)\Delta(k+q)[(k+q)^2-k^2]\,,
\label{pima}
\ee
which, with the notation introduced in  Appendix (\ref{app1}), reads 
\be
I(x) =  \int_{k_{\chic E}} {\widetilde m}^2(y) 
\Delta(y)\Delta(z)\left[1 + \frac{2\sqrt{y}}{\sqrt{x}} \cos\theta \right]\,.
\label{pimb}
\ee
Then, expand $\Delta(z)= \Delta(y) + w \Delta^{\prime}(y) + ...$, and collect the terms that 
survive the angular integration, to obtain  
\bea
I(0) &=&  \int_{k_{\chic E}} {\widetilde m}^2(k^2) 
\Delta(k^2) [\Delta(k^2) + 4 k^2 \Delta^{\prime}(k^2)\cos^2\theta] 
\nonumber\\
&=& \int_{k_{\chic E}} {\widetilde m}^2(k^2) \Delta(k^2) [\Delta(k^2) + k^2 \Delta^{\prime}(k^2)]
\nonumber\\
&=& - \frac{1}{2} \int_{k_{\chic E}} k^2 \Delta^2(k^2) [{\widetilde m}^{2}(k^2)]^{\prime} \,.
\label{pimc} 
\eea
Note that a monotonically decreasing mass,  
$ [{\widetilde m}^{2}(k^2)]^{\prime}<0$,  
guarantees that $I(0)>0$, or, equivalently, the positivity of ${\widetilde m}^{2}(0)$ 
in Euclidean space.  

To write $I(x)$ in a form suitable for solving the corresponding dynamical equation, first
split the radial integration into two intervals, $\int_0^{\infty} dy=  \int_0^{x}dy + \int_x^{\infty}dy$;  
in the first interval apply the usual approximation 
\be
\int_0^{x}dy f_1(z) f_2(y) \approx f_1(x) \int_0^{x} dy f_2(y),
\ee
while in the second, since $x<y$, we can carry out the Taylor expansion as before. Thus, we obtain 
\be
I(x) \approx I_1(x)+ I_2(x)+I_3(x)+I_4(x)\,,
\label{Iterm}
\ee
with
\bea
I_1(x) &=& \Delta(x) \int_0^{x} dy y {\widetilde m}^2(y) \Delta(y) \,,
\nonumber\\
I_2(x) &=&  - \frac{\Delta(x)}{x} \int_0^{x} dy y^2 {\widetilde m}^2(y) \Delta(y) \,,
\nonumber\\
I_3(x) &=& - \frac{1}{2} \,\int_x^{\infty}  dy y^2 \Delta^2(y) [{\widetilde m}^{2}(y)]^{\prime}\,,
\nonumber\\
I_4(x) &=& - \frac{1}{2}\, {\widetilde m}^2(x) \, x^2 \Delta^2(x)\,.
\label{sms} 
\eea
Note that, as $x \to 0$, $I_1(x)$, $I_2(x)$ and $I_4(x)$ vanish, and 
one recovers from $I_3(x)$ the exact result for $I(0)$ given in (\ref{pimc}).  

Finally, 
note that for a $m^2(x)$ displaying the asymptotic behavior given in (\ref{plr})
the following results are useful~\cite{Aguilar:2007ie}, 
\bea
\frac{1}{x}\int_{0}^{x} dy \,m^2(y) &=& \gamma^{-1}   
m^2(x) \ln x + \frac{c^{\prime}}{x} \,,
\nonumber\\ 
\int_{x}^{\infty} dy \,\frac{m^2(y)}{y} &=&   m^2(x) + {\cal O}\left( 1/\ln x\right)\,,
\nonumber\\
\frac{1}{x^2} \int_{0}^{x} dy \,y m^2(y) &=&  m^2(x)  + {\cal O}\left( 1/\ln x\right) \,,
\label{intm23}
\eea
where $c^{\prime}$ is a constant.
The first equation is derived using directly the integral of (\ref{elint}), while for the other two  
we have employed the asymptotic property of the 
incomplete $\Gamma(a,u)$ function. Specifically,   
\be
\Gamma(a,u) = \int_{u}^{\infty} dt\, e^{-t}\, t^{a-1} \,,
\label{Gdef}
\ee
(with no restriction on the sign of $a$),
and its asymptotic representation for large values of $|u|$ is given by
\be
\Gamma(a,u) = u^{a-1} e^{-u} + {\cal O}(|u|^{-1})\,.
\label{Gas}
\ee

\acknowledgments 
\vspace{-0.5cm}

The research of J.~P. is supported by the European FEDER and  Spanish MICINN under grant FPA2008-02878, and the Fundaci\'on General of the UV.


\begin{thebibliography}{99}
\bibitem{Higgs}
In fact, the Higgs mechanism can be viewed as just 
a {\it special realization} of the Schwinger mechanism, 
where the residue of the pole is saturated by $v^2$, 
where  $v$ is the vacuum expectation value of a canonical 
scalar field coupled to the vector meson.
\bibitem{qdi}
In the presence of a hard cutoff 
these integrals are quadratically divergent, up to logarithmic corrections.
\end{thebibliography}
\end{document}